
\let\includefigures=\iftrue
%
\let\useblackboard=\iftrue
%
%
\newfam\black
\input harvmac
\noblackbox 
\includefigures
\message{If you do not have epsf.tex (to include figures),}
\message{change the option at the top of the tex file.}
\input epsf
\def\figin{\epsfcheck\figin}\def\figins{\epsfcheck\figins}
\def\epsfcheck{\ifx\epsfbox\UnDeFiNeD
\message{(NO epsf.tex, FIGURES WILL BE IGNORED)}
\gdef\figin##1{\vskip2in}\gdef\figins##1{\hskip.5in}
\else\message{(FIGURES WILL BE INCLUDED)}%
\gdef\figin##1{##1}\gdef\figins##1{##1}\fi}
\def\DefWarn#1{}
\def\figinsert{\goodbreak\midinsert}
\def\ifig#1#2#3{\DefWarn#1\xdef#1{fig.~\the\figno}
\writedef{#1\leftbracket fig.\noexpand~\the\figno}%
\figinsert\figin{\centerline{#3}}\medskip\centerline{\vbox{
\baselineskip12pt\advance\hsize by -1truein
\noindent\footnotefont{\bf Fig.~\the\figno:} #2}}
\bigskip\endinsert\global\advance\figno by1}
\else
\def\ifig#1#2#3{\xdef#1{fig.~\the\figno}
\writedef{#1\leftbracket fig.\noexpand~\the\figno}%
\global\advance\figno by1} \fi
%

\def\smallfig#1#2#3{\DefWarn#1\xdef#1{fig.~\the\figno}
\writedef{#1\leftbracket fig.\noexpand~\the\figno}%
\figinsert\figin{\centerline{#3}}\medskip\centerline{\vbox{
\baselineskip12pt\advance\hsize by -1truein
\noindent\footnotefont{\bf Fig.~\the\figno:} #2}}
\endinsert\global\advance\figno by1}

\useblackboard
\message{If you do not have msbm (blackboard bold) fonts,}
\message{change the option at the top of the tex file.}
\font\blackboard=msbm10 scaled \magstep1 \font\blackboards=msbm7
\font\blackboardss=msbm5 \textfont\black=\blackboard
\scriptfont\black=\blackboards \scriptscriptfont\black=\blackboardss

\else

\fi
%



\def\boxit#1{\vbox{\hrule\hbox{\vrule\kern8pt
\vbox{\hbox{\kern8pt}\hbox{\vbox{#1}}\hbox{\kern8pt}}
\kern8pt\vrule}\hrule}}
\def\mathboxit#1{\vbox{\hrule\hbox{\vrule\kern8pt\vbox{\kern8pt
\hbox{$\displaystyle #1$}\kern8pt}\kern8pt\vrule}\hrule}}

\def\subsubsection#1{\medskip\noindent
{\it #1}}
\def\subsubsec#1{\subsubsection{#1}}
\def\yboxit#1#2{\vbox{\hrule height #1 \hbox{\vrule width #1
\vbox{#2}\vrule width #1 }\hrule height #1 }}
\def\fillbox#1{\hbox to #1{\vbox to #1{\vfil}\hfil}}
\def\ybox{{\lower 1.3pt \yboxit{0.4pt}{\fillbox{8pt}}\hskip-0.2pt}}
%
%



\def\hg{\hat{g}}

\def\l{\left}

\def\comments#1{}

\def\p{\partial}

\def\eps{\epsilon}
\def\half{{1\over 2}}

\def\bra#1{{\langle}#1|}
\def\ket#1{|#1\rangle}
\def\brket#1#2{\langle #1 | #2 \rangle}

\def\vev#1{\langle{#1}\rangle}

\def\CN{{\cal N}}
\def\CO{{\cal O}}
\def\CP{{\cal P}}
\def\CL{{\cal L}}

\def\CX{{\cal X}}

\def\II{\relax{I\kern-.10em I}}

\font\cmss=cmss10 \font\cmsss=cmss10 at 7pt
\def\IZ{\relax\ifmmode\mathchoice
{\hbox{\cmss Z\kern-.4em Z}}{\hbox{\cmss Z\kern-.4em Z}}
{\lower.9pt\hbox{\cmsss Z\kern-.4em Z}} {\lower1.2pt\hbox{\cmsss
Z\kern-.4em Z}} \else{\cmss Z\kern-.4emZ}\fi}
\def\IR{\relax{\rm I\kern-.18em R}}
\def\IZ{\relax\ifmmode\mathchoice
{\hbox{\cmss Z\kern-.4em Z}}{\hbox{\cmss Z\kern-.4em Z}}
{\lower.9pt\hbox{\cmsss Z\kern-.4em Z}} {\lower1.2pt\hbox{\cmsss
Z\kern-.4em Z}}\else{\cmss Z\kern-.4em Z}\fi}
\def\IB{\relax{\rm I\kern-.18em B}}
\def\IC{{\relax\hbox{$\inbar\kern-.3em{\rm C}$}}}
\def\ID{\relax{\rm I\kern-.18em D}}
\def\IE{\relax{\rm I\kern-.18em E}}
\def\IF{\relax{\rm I\kern-.18em F}}
\def\IG{\relax\hbox{$\inbar\kern-.3em{\rm G}$}}
\def\IGa{\relax\hbox{${\rm I}\kern-.18em\Gamma$}}
\def\IH{\relax{\rm I\kern-.18em H}}
\def\II{\relax{\rm I\kern-.18em I}}
\def\IK{\relax{\rm I\kern-.18em K}}
\def\IP{\relax{\rm I\kern-.18em P}}

%

\def\inbar{\,\vrule height1.5ex width.4pt depth0pt}

\def\p{\partial}

\font\cmss=cmss10 
\def\IR{\relax{\rm I\kern-.18em R}}

%


%

\def\lp10{\ell_p^{10}}
\def\lp11{\ell_p^{11}}
\def\R11{R_{11}}

\def\frac#1#2{{#1 \over #2}}



\def\l{\left}

\def\comments#1{}

\def\p{\partial}

\def\eps{\epsilon}
\def\half{{1\over 2}}

\def\bra#1{{\langle}#1|}
\def\ket#1{|#1\rangle}

\def\vev#1{\langle{#1}\rangle}

\def\CH{{\cal H}}

\def\CN{{\cal N}}
\def\CO{{\cal O}}
\def\CP{{\cal P}}
\def\CL{{\cal L}}

\def\CX{{\cal X}}


\def\tg{{\tilde g}}

\def\csch{{\rm\ csch\ }}


\def\cf{{\it c.f.}}

\def\M4{M_{Pl,4}}

\def\k11{\kappa_{11}}
\def\l11{\ell_{11}}
\def\tl11{\tilde{\ell}_{11}}

\def\m11{M_{11}}
\def\tm11{\tilde{M}_{11}}

\def\etal{{\it et.\ al.}}
\def\eg{{\it e.g.}}

\def\({\left(}
\def\){\right)}
\def\[{\left[}
\def\]{\right]}
\def\<{\langle}
\def\>{\rangle}
\def\half{{1\over 2}}
\def\d{\partial}

\def\|{\biggl|}

\def\bx{{\bf x}}
\def\by{{\bf y}}
\def\ba{{\bf a}}
\def\bb{{\bf b}}

\lref\FreedmanWX{
  D.~Z.~Freedman, M.~Headrick and A.~Lawrence,
  ``On closed string tachyon dynamics,''
  arXiv:hep-th/0510126.
}

\lref\AkhmedovVF{
  E.~T.~Akhmedov,
  ``A remark on the AdS/CFT correspondence and the renormalization group
  flow,''
  Phys.\ Lett.\ B {\bf 442}, 152 (1998)
  [arXiv:hep-th/9806217].
}
\lref\BalasubramanianJD{
  V.~Balasubramanian and P.~Kraus,
  ``Spacetime and the holographic renormalization group,''
  Phys.\ Rev.\ Lett.\  {\bf 83}, 3605 (1999)
  [arXiv:hep-th/9903190].
}
\lref\deBoerXF{
  J.~de Boer, E.~P.~Verlinde and H.~L.~Verlinde,
  ``On the holographic renormalization group,''
  JHEP {\bf 0008}, 003 (2000)
  [arXiv:hep-th/9912012].
}
\lref\deBoerCZ{
  J.~de Boer,
  ``The holographic renormalization group,''
  Fortsch.\ Phys.\  {\bf 49}, 339 (2001)
  [arXiv:hep-th/0101026].
}
\lref\KalkkinenVG{
  J.~Kalkkinen, D.~Martelli and W.~Muck,
  ``Holographic renormalisation and anomalies,''
  JHEP {\bf 0104}, 036 (2001)
  [arXiv:hep-th/0103111].
}
\lref\PorratiEW{
  M.~Porrati and A.~Starinets,
  ``RG fixed points in supergravity duals of 4-d field theory and
  asymptotically AdS spaces,''
  Phys.\ Lett.\ B {\bf 454}, 77 (1999)
  [arXiv:hep-th/9903085].
}
\lref\AlvarezWR{
  E.~Alvarez and C.~Gomez,
  ``Geometric holography, the renormalization group and the c-theorem,''
  Nucl.\ Phys.\ B {\bf 541}, 441 (1999)
  [arXiv:hep-th/9807226].
}
\lref\SkenderisWP{
  K.~Skenderis,
  ``Lecture notes on holographic renormalization,''
  Class.\ Quant.\ Grav.\  {\bf 19}, 5849 (2002)
  [arXiv:hep-th/0209067].
}

\lref\EvansEQ{
  N.~Evans, T.~R.~Morris and O.~J.~Rosten,
  ``Gauge invariant regularization in the AdS/CFT correspondence and ghost
  D-branes,''
  arXiv:hep-th/0601114.
}
\lref\HellermanQA{
  S.~Hellerman,
  ``Lattice gauge theories have gravitational duals,''
  arXiv:hep-th/0207226.
}
\lref\SkenderisUY{
  K.~Skenderis and M.~Taylor,
  ``Kaluza-Klein holography,''
  arXiv:hep-th/0603016.
}

\lref\FukumaBZ{
  M.~Fukuma, S.~Matsuura and T.~Sakai,
  ``A note on the Weyl anomaly in the holographic renormalization group,''
  Prog.\ Theor.\ Phys.\  {\bf 104}, 1089 (2000)
  [arXiv:hep-th/0007062].
}
\lref\HenningsonGX{
  M.~Henningson and K.~Skenderis,
  ``The holographic Weyl anomaly,''
  JHEP {\bf 9807}, 023 (1998)
  [arXiv:hep-th/9806087].
}
\lref\ErdmengerJA{
  J.~Erdmenger,
  ``A field-theoretical interpretation of the holographic renormalization
  group,''
  Phys.\ Rev.\ D {\bf 64}, 085012 (2001)
  [arXiv:hep-th/0103219].
}

\lref\BalasubramanianSN{
  V.~Balasubramanian, P.~Kraus and A.~E.~Lawrence,
  ``Bulk vs. boundary dynamics in anti-de Sitter spacetime,''
  Phys.\ Rev.\ D {\bf 59}, 046003 (1999)
  [arXiv:hep-th/9805171].
}
\lref\BalasubramanianDE{
  V.~Balasubramanian, P.~Kraus, A.~E.~Lawrence and S.~P.~Trivedi,
  ``Holographic probes of anti-de Sitter space-times,''
  Phys.\ Rev.\ D {\bf 59}, 104021 (1999)
  [arXiv:hep-th/9808017].
}
\lref\BanksDD{
  T.~Banks, M.~R.~Douglas, G.~T.~Horowitz and E.~J.~Martinec,
  ``AdS dynamics from conformal field theory,''
  arXiv:hep-th/9808016.
}
\lref\BalasubramanianRI{
  V.~Balasubramanian, S.~B.~Giddings and A.~E.~Lawrence,
  ``What do CFTs tell us about anti-de Sitter spacetimes?,''
  JHEP {\bf 9903}, 001 (1999)
  [arXiv:hep-th/9902052].
}
\lref\MarolfFY{
  D.~Marolf,
  ``States and boundary terms: Subtleties of Lorentzian AdS/CFT,''
  JHEP {\bf 0505}, 042 (2005)
  [arXiv:hep-th/0412032].
}

\lref\MaldacenaRE{
  J.~M.~Maldacena,
  ``The large N limit of superconformal field theories and supergravity,''
  Adv.\ Theor.\ Math.\ Phys.\  {\bf 2}, 231 (1998)
  [Int.\ J.\ Theor.\ Phys.\  {\bf 38}, 1113 (1999)]
  [arXiv:hep-th/9711200].
}
\lref\GubserBC{
  S.~S.~Gubser, I.~R.~Klebanov and A.~M.~Polyakov,
  ``Gauge theory correlators from non-critical string theory,''
  Phys.\ Lett.\ B {\bf 428}, 105 (1998)
  [arXiv:hep-th/9802109].
}
\lref\WittenQJ{
  E.~Witten,
  ``Anti-de Sitter space and holography,''
  Adv.\ Theor.\ Math.\ Phys.\  {\bf 2}, 253 (1998)
  [arXiv:hep-th/9802150].
}
\lref\FreedmanTZ{
  D.~Z.~Freedman, S.~D.~Mathur, A.~Matusis and L.~Rastelli,
  ``Correlation functions in the CFT($d$)/AdS($d+1$) correspondence,''
  Nucl.\ Phys.\ B {\bf 546}, 96 (1999)
  [arXiv:hep-th/9804058].
}
\lref\AharonyTI{
  O.~Aharony, S.~S.~Gubser, J.~M.~Maldacena, H.~Ooguri and Y.~Oz,
  ``Large N field theories, string theory and gravity,''
  Phys.\ Rept.\  {\bf 323}, 183 (2000)
  [arXiv:hep-th/9905111].
}
\lref\PolchinskiYD{
  J.~Polchinski, L.~Susskind and N.~Toumbas,
  ``Negative energy, superluminosity and holography,''
  Phys.\ Rev.\ D {\bf 60}, 084006 (1999)
  [arXiv:hep-th/9903228].
}
\lref\SusskindEY{
  L.~Susskind and N.~Toumbas,
  ``Wilson loops as precursors,''
  Phys.\ Rev.\ D {\bf 61}, 044001 (2000)
  [arXiv:hep-th/9909013].
}

\lref\SusskindDQ{
  L.~Susskind and E.~Witten,
  ``The holographic bound in anti-de Sitter space,''
  arXiv:hep-th/9805114.
}
\lref\BanksNR{
  T.~Banks and M.~B.~Green,
  ``Non-perturbative effects in AdS(5) x S**5 string theory and d = 4 SUSY
  Yang-Mills,''
  JHEP {\bf 9805}, 002 (1998)
  [arXiv:hep-th/9804170].
}
\lref\ChuIN{
  C.~S.~Chu, P.~M.~Ho and Y.~Y.~Wu,
  ``D-instanton in AdS(5) and instanton in SYM(4),''
  Nucl.\ Phys.\ B {\bf 541}, 179 (1999)
  [arXiv:hep-th/9806103].
}
\lref\KoganRE{
  I.~I.~Kogan and G.~Luzon,
  ``D-instantons on the boundary,''
  Nucl.\ Phys.\ B {\bf 539}, 121 (1999)
  [arXiv:hep-th/9806197].
}
\lref\BianchiNK{
  M.~Bianchi, M.~B.~Green, S.~Kovacs and G.~Rossi,
  ``Instantons in supersymmetric Yang-Mills and D-instantons in IIB
  superstring theory,''
  JHEP {\bf 9808}, 013 (1998)
  [arXiv:hep-th/9807033].
}
\lref\PeetWN{
  A.~W.~Peet and J.~Polchinski,
  ``UV/IR relations in AdS dynamics,''
  Phys.\ Rev.\ D {\bf 59}, 065011 (1999)
  [arXiv:hep-th/9809022].
}
\lref\ZamolodchikovTI{
  A.~B.~Zamolodchikov,
  ``Renormalization Group And Perturbation Theory Near Fixed Points In
  Two-Dimensional Field Theory,''
  Sov.\ J.\ Nucl.\ Phys.\  {\bf 46}, 1090 (1987)
  [Yad.\ Fiz.\  {\bf 46}, 1819 (1987)].
}
\lref\CardyXT{
  J.~Cardy,
  {\it Scaling and renormalization in statistical physics}, Cambridge (1996)
}
%
\lref\AmitMS{
  D.~J.~Amit,
  {\it Field Theory, The Renormalization Group, And Critical Phenomena,'}
  World Scientific (1984), 2nd ed.
}

\lref\ReyIK{
  S.~J.~Rey and J.~T.~Yee,
  ``Macroscopic strings as heavy quarks in large N gauge theory and  anti-de
  Sitter supergravity,''
  Eur.\ Phys.\ J.\ C {\bf 22}, 379 (2001)
  [arXiv:hep-th/9803001].
}

\lref\MaldacenaIM{
  J.~M.~Maldacena,
  ``Wilson loops in large N field theories,''
  Phys.\ Rev.\ Lett.\  {\bf 80}, 4859 (1998)
  [arXiv:hep-th/9803002].
}

\lref\OsbornGM{
  H.~Osborn,
  ``Weyl consistency conditions and a local renormalization group equation for
  general renormalizable field theories,''
  Nucl.\ Phys.\ B {\bf 363}, 486 (1991).
}

\lref\AharonyPA{
  O.~Aharony, M.~Berkooz and E.~Silverstein,
  ``Multiple-trace operators and non-local string theories,''
  JHEP {\bf 0108}, 006 (2001)
  [arXiv:hep-th/0105309].
}
\lref\WittenUA{
  E.~Witten,
  ``Multi-trace operators, boundary conditions, and AdS/CFT correspondence,''
  arXiv:hep-th/0112258.
}
\lref\BerkoozUG{
  M.~Berkooz, A.~Sever and A.~Shomer,
  ``Double-trace deformations, boundary conditions and spacetime
  singularities,''
  JHEP {\bf 0205}, 034 (2002)
  [arXiv:hep-th/0112264].
}
\lref\SeverFK{
  A.~Sever and A.~Shomer,
  ``A note on multi-trace deformations and AdS/CFT,''
  JHEP {\bf 0207}, 027 (2002)
  [arXiv:hep-th/0203168].
}

\lref\BerryJV{
  M.~V.~Berry,
  ``Quantal Phase Factors Accompanying Adiabatic Changes,''
  Proc.\ Roy.\ Soc.\ Lond.\ A {\bf 392}, 45 (1984).
}
\lref\SimonMH{
  B.~Simon,
  ``Holonomy, The Quantum Adiabatic Theorem, And Berry's Phase,''
  Phys.\ Rev.\ Lett.\  {\bf 51}, 2167 (1983).
}
\lref\WilczekDH{
  F.~Wilczek and A.~Zee,
  ``Appearance Of Gauge Structure In Simple Dynamical Systems,''
  Phys.\ Rev.\ Lett.\  {\bf 52}, 2111 (1984).
}
\lref\ShapereKP{
  A.~D.~Shapere and F.~Wilczek, (eds),
  ``Geometric Phases In Physics,''
  Adv.\ Ser.\ Math.\ Phys.\  {\bf 5}, 1 (1989).
}

\lref\YaffeVF{
  L.~G.~Yaffe,
  ``Large N Limits As Classical Mechanics,''
  Rev.\ Mod.\ Phys.\  {\bf 54}, 407 (1982).
}
\lref\HellermanQA{
  S.~Hellerman,
  ``Lattice gauge theories have gravitational duals,''
  arXiv:hep-th/0207226.
}

\lref\WilsonZS{
  K.~G.~Wilson,
  ``Nonlagrangian Models Of Current Algebra,''
  Phys.\ Rev.\  {\bf 179}, 1499 (1969).
}

\lref\JackiwKV{
  R.~Jackiw and S.~Templeton,
  ``How Superrenormalizable Interactions Cure Their Infrared Divergences,''
  Phys.\ Rev.\ D {\bf 23}, 2291 (1981).
}
\lref\ZamolodchikovBK{
  A.~B.~Zamolodchikov,
  ``Two Point Correlation Function In Scaling Lee-Yang Model,''
  Nucl.\ Phys.\ B {\bf 348}, 619 (1991).
}
\lref\BelavinPU{
  A.~A.~Belavin, V.~A.~Belavin, A.~V.~Litvinov, Y.~P.~Pugai and A.~B.~Zamolodchikov,
  ``On correlation functions in the perturbed minimal models M(2,2n+1),''
  Nucl.\ Phys.\ B {\bf 676}, 587 (2004)
  [arXiv:hep-th/0309137].
}
\lref\BerensteinJQ{
  D.~Berenstein, J.~M.~Maldacena and H.~Nastase,
  ``Strings in flat space and pp waves from N = 4 super Yang Mills,''
  JHEP {\bf 0204}, 013 (2002)
  [arXiv:hep-th/0202021].
}
\lref\AtickSI{
  J.~J.~Atick and E.~Witten,
  ``The Hagedorn Transition And The Number Of Degrees Of Freedom Of String
  Theory,''
  Nucl.\ Phys.\ B {\bf 310}, 291 (1988).
}

\lref\BreitenlohnerBM{
  P.~Breitenlohner and D.~Z.~Freedman,
  ``Positive Energy In Anti-De Sitter Backgrounds And Gauged Extended
  Supergravity,''
  Phys.\ Lett.\ B {\bf 115}, 197 (1982).
}
\lref\BreitenlohnerJF{
  P.~Breitenlohner and D.~Z.~Freedman,
  ``Stability In Gauged Extended Supergravity,''
  Annals Phys.\  {\bf 144}, 249 (1982).
}

\lref\KlebanovTB{
  I.~R.~Klebanov and E.~Witten,
  ``AdS/CFT correspondence and symmetry breaking,''
  Nucl.\ Phys.\ B {\bf 556}, 89 (1999)
  [arXiv:hep-th/9905104].
}

\lref\HartmanDY{
  T.~Hartman and L.~Rastelli,
  ``Double-trace deformations, mixed boundary conditions and functional
  determinants in AdS/CFT,''
  arXiv:hep-th/0602106.
}

\lref\WittenZW{
  E.~Witten,
  ``Anti-de Sitter space, thermal phase transition, and confinement in  gauge
  theories,''
  Adv.\ Theor.\ Math.\ Phys.\  {\bf 2}, 505 (1998)
  [arXiv:hep-th/9803131].
}
\lref\HorowitzBJ{
  G.~T.~Horowitz and H.~Ooguri,
  ``Spectrum of large N gauge theory from supergravity,''
  Phys.\ Rev.\ Lett.\  {\bf 80}, 4116 (1998)
  [arXiv:hep-th/9802116].
}
\lref\PolchinskiUF{
  J.~Polchinski and M.~J.~Strassler,
  ``The string dual of a confining four-dimensional gauge theory,''
  arXiv:hep-th/0003136.
}
\lref\RandallEE{
  L.~Randall and R.~Sundrum,
  ``A large mass hierarchy from a small extra dimension,''
  Phys.\ Rev.\ Lett.\  {\bf 83}, 3370 (1999)
  [arXiv:hep-ph/9905221].
}

\lref\PolchinskiUF{
  J.~Polchinski and M.~J.~Strassler,
  arXiv:hep-th/0003136.
}

\lref\BergHY{
  M.~Berg and H.~Samtleben,
   ``Holographic correlators in a flow to a fixed point,''
  JHEP {\bf 0212}, 070 (2002)
  [arXiv:hep-th/0209191].
}

\lref\TreimanEP{
  S.~B.~Treiman, E.~Witten, R.~Jackiw and B.~Zumino,
}

\Title{\vbox{\baselineskip12pt\hbox{hep-th/0606022} \hbox{BRX
TH-574}}} {\vbox{ \centerline{Holography and renormalization}
\smallskip
\smallskip
\smallskip
\centerline{in Lorentzian signature} }} \centerline{Albion Lawrence
and Amit Sever}
\bigskip
\centerline{{Brandeis Theory Group, Martin Fisher School of
Physics,}} \centerline{{Brandeis University, Waltham, MA
02454-9110}}

\bigskip
\bigskip
\noindent

De Boer {\it et. al.} have found an asymptotic equivalence between
the Hamilton-Jacobi equations for supergravity in
$(d+1)$-dimensional asymptotic anti-de Sitter space, and the
Callan-Symanzik equations for the dual $d$-dimensional perturbed conformal
field theory. We discuss this correspondence in Lorentzian
signature. We construct a gravitational
dual of the generating function of correlation functions between
initial and final states, in accordance with the construction of Marolf,
and find a class of states for which the
result has a classical
supergravity limit. We show how the data specifying the full set of
solutions to the second-order supergravity equations of motion are
described in the field theory, despite the first-order nature of the
renormalization group equations for the running couplings: one must
specify both the couplings and the states, and the latter affects
the solutions to the Callan-Symanzik equations.

\Date{June 2006}

\listtoc
\writetoc

\newsec{Introduction}

In the AdS${}_{d+1}$/CFT${}_d$ correspondence \refs{\MaldacenaRE},
the coordinate position of an excitation relative to the timelike
boundary of $AdS$ is in some sense dual to the characteristic
scale size of that excitation in the $d$-dimensional CFT.  This can be seen from entropic
considerations \refs{\SusskindDQ}, from the duals of classical bulk
probes
\refs{\BanksNR\ChuIN\KoganRE\BianchiNK\BalasubramanianDE-\PeetWN},
and from the semiclassical bulk description of Wilson lines
\refs{\ReyIK,\MaldacenaIM}.

Let ${\bf x}$ be coordinates parallel to the boundary and $r$ the
coordinate running perpendicular to the boundary. In the Euclidean
version of the correspondence, the bulk fields $\phi^a({\bf x},r)$
are taken to be dual to the (in general space-time dependent)
couplings $\lambda^a({\bf x})$ of the boundary theory.\foot{Here
${\bf x}$ parameterizes a point on the $d$-dimensional boundary.}
The equations of motion for $\phi$ can be written as an equation for
evolution in $r$, where the boundary of AdS space is dual to the UV
of the field theory.  The "holographic renormalization group"
\refs{\AkhmedovVF\AlvarezWR\PorratiEW\BalasubramanianJD\deBoerXF-\deBoerCZ}
then relates this evolution to the running of the dual couplings
under change of renormalization group scale.

In particular, de Boer, Verlinde, and Verlinde have shown that as we
approach the AdS boundary, the Hamilton-Jacobi (HJ) equations for
radial evolution of the bulk supergravity fields are equivalent to
Callan-Symanzik (CS) equations for the correlation functions of the
boundary field theory \refs{\deBoerXF,\deBoerCZ}. In this formalism,
all nonsingular solutions to the supergravity equations (such was
domain walls in AdS), and solutions with singularities that are
resolved by string theory, are manifestly dual to renormalization
group flows.

Nonetheless, this and other versions of holographic RG raise a
number of issues, of which we list three:

\smallskip
\smallskip
\item{\bf 1.} The CS equations are first order in the RG scale parameter,
while the spacetime equations of motion are second order.  In
Lorentzian signature, the spacetime equations of motion seem to
require twice as many initial conditions
\refs{\BalasubramanianDE,\BalasubramanianSN}\ than are typically
specified in renormalization group flows.

\item{\bf 2.} As we approach the AdS boundary, $\phi(r)$
in general becomes the coupling of the dual operator at scale
$\ell(r)$ (for low-dimension operators this does not have to be true
\refs{\BalasubramanianDE,\KlebanovTB}.) This is not obviously the
right map deep in the interior of $AdS$ spacetimes
\refs{\KalkkinenVG,\BergHY}. Furthermore, in Lorentzian signature
\refs{\BalasubramanianDE,\BalasubramanianSN}, $\phi$ is determined
by both the coupling and the state of the dual field theory.

\item{\bf 3.} Considered as flow equations in the radial direction $r$ of
AdS, the spacetime equations of motion are reversible. The
Callan-Symanzik equations are also reversible in RG scale. On the
other hand, the Wilsonian version of the renormalization group is an
evolution under coarse graining, which is not reversible. How then does
the Wilsonian picture fit into the AdS/CFT correspondence?

\vskip .2cm

The main goal of this paper is to solve the puzzle raised in the
first point. The summary of the solution is as follows.  In
Lorentzian signature, the second order supergravity equations of
motion have two classes of nonsingular solutions characterized by
their behavior at the timelike AdS boundary
\refs{\BalasubramanianSN,\BalasubramanianDE}. One class is dual to
deformations of the field theory Lagrangian.  The second class
depends on the semiclassical excitations of the field theory above
the vacuum state.\foot{In this class there are also singular
solutions in either signature generated by sources such as D-branes or
D-instantons.} A general solution to the spacetime equations
of motion will have terms with both types of boundary behavior, and
so be specified both the by couplings of the perturbed CFT, and by
the state of the CFT, when the state is a "classical" state in the
large-N limit.

In the Hamilton-Jacobi formulation, the equations of motion are
solved by first solving the Hamilton-Jacobi equations for Hamilton's
principal function $S$:
$$ H(\phi^a, \pi_{a,\phi} = \frac{\p S}{\p\phi^a}) = 0 $$\
and then solving Hamilton's equations
$$ \dot{\phi}^a = \frac{\p H}{\p\pi_{a,\phi}}(\pi_a =
   \frac{\p S}{\p\phi^a})\ , $$
where $\phi^a$ are some fields in AdS, $\pi_{a,\phi}$ are the
conjugate field momenta and the dot stands for radial derivative. In
holographic renormalization the former equations are dual to the
Callan-Symanzik equations, while the latter are dual to the RG
equations
$$ \Lambda \p_{\Lambda} \lambda^a = \beta^a(\lambda) $$
for the couplings $\lambda$, where $\beta^a$ are the beta functions
and $\Lambda$ is a momentum space cutoff. We will show that in these
solutions, the freedom to excite modes dual to a choice of state is
captured in the choice of solutions to the Hamilton-Jacobi equation.
For the holographic equivalence between the Hamilton-Jacobi and
Callan-Symanzik equations to hold, we must be able to solve the
latter for any choice of state. We find that we can, once we take
into account the modification of the Callan-Symanzik equation for
correlation functions in a nonvacuum state.

Along the way we will also discuss the problem raised in (2), and
resolve the problem raised in (3).

The format of this paper is as follows.  In \S2\ we review HJ theory
and apply it to some simple examples, pointing out specific features
useful for our discussion. In \S3\ following
\refs{\BalasubramanianDE,\BalasubramanianSN,\MarolfFY}, we discuss
the AdS/CFT correspondence in Lorentzian signature. In \S3.1\ we review the AdS/CFT
basics. In \S3.2\ the boundary behavior of classical fields in AdS
is discussed. In \S3.3\ we explain the classical supergravity
manifestation of the CFT states. We find that in order to guarantee
the existence of a saddle point over a range of couplings, the eigenstates of the field
operators are good choices for initial and final states.  In \S3.4\
we discuss the issue of gravitational backreaction for such states.
In \S3.5\ we review the correspondence where the CFT is perturbed by
relevant operators. Finally in \S3.6\ we discusses the generating
function of correlation functions. \S4\ is a review and critical
discussion of the formalism of \refs{\deBoerXF,\deBoerCZ}. In \S5\
we develop a Lorentzian-signature version of the picture in
\refs{\deBoerXF,\deBoerCZ}. We identify the "missing constants of
motion" in the RG equations with the choice of classical state of
the system. In \S5.1\ we discuss the Hamilton-Jacobi equations in
Lorentzian signature. In \S5.2\ we derive a Callan-Symanzik equation
for general matrix elements of time-ordered products of operators.
In \S5.3\ we show that the Hamilton-Jacobi and Callan-Symanzik
equations are determined by the same information, thus solving the
puzzle posed in question (1). In \S5.4\ we discuss an alternate
solution to the Hamilton-Jacobi equations, in which the constants of
motion are the operator expectation values specified at an
infrared cutoff. In \S5.5\ we discuss the extension of our story
deep into the infrared. In \S5.6\ we discuss the degree to which
holographic RG is related to the Wilsonian picture of
renormalization.  In \S6\ we conclude.

\newsec{Review of Hamilton-Jacobi theory}

Consider a dynamical system with $2n$ phase space variables $({\bf
q},{\bf p})$, corresponding to positions ${\bf q}=\{q_i,\ i=1\dots
n\}$ and canonical conjugate momenta ${\bf p}=\{p_i,\ i=1\dots n\}$,
and a Hamiltonian $H({\bf p},{\bf q})$.  In Hamilton-Jacobi theory,
the equations of motion are solved in two stages.  First, one solves
the Hamilton-Jacobi equation for Hamilton's principal function $S$:
\eqn\hjequation{
    \p_t S({\bf q},t) + H\left({\bf p}=\frac{\p S({\bf q},t)}{\p {\bf q}}, {\bf q}\right) = 0~.
}
This is a nonlinear equation; in general it has many solutions.
Given a solution, one finds the classical trajectories ${\bf q}(t)$
from a set of first-order differential equations:
\eqn\hjdiffeqs{
    {\bf \dot q}=\frac{\p H}{\p {\bf p}}\biggl|_{{\bf p}=\frac{\p S}{\p
    {\bf q}}}~.
}

If $H$ is quadratic in momenta, the full equations of motion for $q$
are second order in time.  Their full soutions require that one
specify $2n$ constants of motion $({\bf a},{\bf b})$, where ${\bf
a}=\{a_i,\ i=1\dots n\}$ and ${\bf b}=\{b_i,\ i=1\dots n\}$. For
example, ${\bf q}(t_i)$ and $\dot {\bf q}(t_i)$ at some initial time
$t_i$ determine the trajectory completely. On the other hand, a full
solution to \hjdiffeqs\ requires only $n$ constants of motion (${\bf
b}$), for example ${\bf b}={\bf q}(t_i)$ at some initial time $t_i$.

The point is that the additional constants of motion of the dynamics
are contained in the choice of solution to \hjequation. The solution
can be written as $S({\bf q}(t),{\bf a},t)$, where ${\bf a}$ are $n$
constants of motion. If in addition to \hjdiffeqs\ we demand that
 \eqn\demand{\|{\d^2S({\bf q}(t),{\bf a};t)\over\d q^i\d a^j}\|\ne 0~,}
then the constants of motion ${\bf b}$ are given by
 \eqn\bconst{{\bf b}=-{\d S({\bf q}(t),{\bf a},t)\over\d{\bf a}}~,}
${\bf b}$ is canonically conjugate to ${\bf a}$ and $S$ is the
generating function of canonical transformation between $({\bf
q},{\bf p})$ and $({\bf a},{\bf b})$. Since the new canonical
variables (${\bf a},{\bf b}$) are constants of motion, the new
Hamiltonian
\eqn\Hamilton{K({\bf a},{\bf b},t)= \p_t S + H = 0~.}
Note that, instead of using \hjdiffeqs\ one can extract the solution
${\bf q}({\bf a},{\bf b},t)$ directly from \bconst.

One particular choice of ${\bf a}$ is ${\bf q}(t_0)$ at some initial
time $t_0$. The corresponding solution to \hjequation\ is:
\eqn\hjaction{
    S({\bf q}(t),{\bf q}(t_0),t) = \int^t_{t_0} dt' \left( {\bf p\dot q} - H \right)~,}
evaluated on a solution to the classical equations of motion with
fixed ${\bf q}(t_0)\equiv {\bf q}_0 = {\bf a}$.

\subsec{Example: the upside-down harmonic oscillator}

As an example, let us study the one-dimensional upside-down harmonic
oscillator, with Hamiltonian $H = \half p^2 - \half \Omega^2 q^2$.
Since $H$ is time independent, one solution to \hjequation\ can be
found by setting $S_1 = - E t + W(q)$. The HJ equation becomes a
differential equation for $W$:
\eqn\hprinceq{
    \half\[(\p_q W)^2 - \Omega^2 q^2\] = E~,
}
which has the solution:
\eqn\princeqsolution{
    W(q,E)  = {E\over\Omega}\sinh^{-1}\({\Omega q\over \sqrt{2E}}\)
        +{\Omega \over 2}q\sqrt{{2E\over\Omega^2} + q^2} + f_1(E)~.
}
Here ${\bf a} = E$ is the constant of motion governing the solution
to the Hamilton-Jacobi equation. $f_1(E)$ is an arbitrary function;
$f_1$ changes the definition of the phase space variable conjugate
to $E$. The equation of motion for $q$ now reduces to:
\eqn\firsteom{
    \dot q = p = \p_q W = \sqrt{2 E + \Omega^2 q^2}~,
}
and has the solution
\eqn\firstsolution{
    q(t) = \frac{\sqrt{2E}}{\Omega} \sinh\[\Omega(t - t_0)\]~.}
Here
\eqn\integconstant{
    {\bf b} \equiv -{\d S_1\over\d E}=t_0 - f_1'(E)
}
is the integration constant arising from the first order
differential equation \firsteom. The complete solution is specified
by $t_0$ and $E$, where $t_0$ is defined as the time at which
$q(t_0) = 0$.  The solution $S_1$ is the generating function of the
canonical transformation between $(q,p)$ and $(E,t_0-f_1'(E))$.

Alternatively, we can find the classical action for $q(t)$ given
that $q(t_0) = q_0$. Here ${\bf a} = q_0$ is the constant of motion
that arises in the solution to \hjequation:
\eqn\initcondsoln{
    S_2(q,q_0,t) = \half\Omega\coth\[\Omega (t-t_0)\](q^2 + q_0^2) -
    \Omega\csch\[\Omega (t-t_0)\]qq_0 + f_2(q_0)~.
}
where $f_2$ is an arbitrary function, changing the definition of the
momentum conjugate to $q_0$. Eq. \initcondsoln\ can be computed by
simply inserting the known classical solution into the classical
action $S = \int_{t_0}^t L$. The equation of motion for $q$ is:
\eqn\secondeom{
    \dot q = \p_q S_2 = \Omega\coth\[\Omega(t - t_0)\]q - \Omega\csch\[\Omega(t -
    t_0)\]q_0~.
}
Integrating this, we find that
\eqn\secondsoln{
    q(t) = q_0\cosh\[\Omega(t -t_0)\]+\frac{p_0}{\Omega} \sinh\[\Omega (t -
    t_0)\]~.
}
Here
\eqn\intconst{\bb\equiv -{\d S_2\over\d q_0}=p_0-f_2'(q_0)}
is the integration constant arising from the first order equation
\secondeom.  $S_2$ is the generating function of the canonical
transformation between $(q,p)$ and $(q_0,p_0-f_2'(q_0))$.

We have solved for the dynamics by first choosing the constants of
motion ${\bf a}$ in the solution to the Hamilton-Jacobi equation,
and then solving for the trajectory $q(t)$ via Hamilton's equations.
Note, however, that for a given ${\bf a}$, not all values of $q$ may
be allowed. For example, consider trajectories with fixed energy for
the standard harmonic oscillator with frequency $\omega$. The
solution can be found from \princeqsolution\ by setting $\Omega =
i\omega$. For fixed $\omega$, the region $q > 2E/(m\omega^2)$ is
classically forbidden.  This appears already at the level of the
solution $W$, which becomes imaginary in this region.  While $W$ in
this region can be used in a WKB analysis (as the phase of the WKB
wavefunction satisfies the Hamilton-Jacobi equation to lowest order
in $\hbar$), it does not correspond to any classical trajectory.

\newsec{AdS/CFT in Lorentzian signature}

We will be discussing the gravitational duals of perturbed
$d$-dimensional conformal field theories. Consider a CFT $\CX$
perturbed by spacetime dependent couplings $\lambda^a(x)$ to local
operators $\CO_a(x)$. Correlation functions of local operators can
be extracted from the transition amplitudes of the perturbed theory:
\eqn\transamp{
    Z[\{\lambda(x)\}] =
    \bra{\psi_+(t_+)} T \exp\left(\frac{- i}{\hbar} \int_{t = t_-}^{t = t_+}
    d^d x \sum_a \lambda^a(x) \CO_a(x)\right)
        \ket{\psi_-(t_-)}~,
}
by taking functional derivatives with respect to $\lambda^a(x)$.
Here the local operators $\CO_a(x)$ and the states
$\ket{\psi_\pm}$ are written in the interaction picture.

The generating function of Euclidean correlators was constructed in
\refs{\GubserBC,\WittenQJ}. This study of the duality in Lorentzian
signature was initiated in
\refs{\BalasubramanianDE,\BalasubramanianSN}, which provided a
duality map for the propagating classical and quantum states. In
\MarolfFY\ these states were constructed in the bulk, so that they
are independent of variations of $\lambda$ in the interior of $[t_-,t_+]$.
This allows \transamp\ to be the
generating function of matrix elements of time-ordered products of
operators.

This section will be dedicated to sketching the gravitational dual
of \transamp\ in the semiclassical limit, following
\refs{\BalasubramanianDE,\BalasubramanianSN,\MarolfFY}.  Following
\refs{\MarolfFY}, we take care to define $\ket{\psi_{\pm}}$ so that
they are independent of the coupling.  In particular, in later
sections we will be interested in applying this formula in a
classical limit where $Z$ can be considered as a solution to the
Hamilton-Jacobi equation of the dual supergravity theory, with
$\lambda^a(x)$ as the configuration space variables. We can apply
classical Hamilton-Jacobi theory to \transamp\ if there is a
solution to the classical supergravity equations for each value of
$\lambda(x)$.  We will argue that for these purposes, choosing
$\ket{\psi_\pm}$ to be eigenstates of the field operators in the
gravitational dual will lead to $Z$ having the desired properties,
and are a technically convenient choice.
These states are potentially dangerous in a theory of quantum
gravity. We will discuss these dangers and the reasons why they
should not trouble the semiclassical computation of \transamp.

We will take the CFT to be
$d=4$, $\CN=4$ supersymmetric Yang-Mills theory with gauge group
$U(N)$. The story for other CFTs will be essentially the same,
even in other spacetime dimensions.

\subsec{AdS/CFT basics}

We begin by reviewing general aspects of this duality, as outlined
by \refs{\MaldacenaRE,\GubserBC,\WittenQJ}. The correspondence can
be considered in various coordinate patches of AdS spacetime. A set
of coordinates which cover the entire Lorentzian spacetime are:
\eqn\globalcoords{
    ds^2 = - \left(1 + \frac{r^2}{R_{AdS}^2}\right) dt^2  + \frac{dr^2}{\left(1 + \frac{r^2}{R_{AdS}^2}\right)}
        + r^2 d\Omega^2_3~,
}
where $d\Omega_3$ is the solid angle on $S^3$, and $R_{AdS}$ is the
$AdS$ radius of curvature. There is a timelike boundary at $r =
\infty$ which is conformally equivalent to $\IR_t \times S^3$.  This
boundary is at infinite proper distance along $r$, but light rays
can reach the boundary in finite global time (see Fig. 1.) String
theory on the full $AdS_5\times S^5$ is dual to $d=4$, $\CN=4$
super-Yang-Mills theory on $\IR_t\times S^3$.

\ifig\angles{A Penrose diagram for $AdS_{d+1}$.  The cylindrical
boundary conformal to $S^{d-1}\times\IR$. $t$ is the time in global
coordinates.  $\Sigma_{\pm}$ are spacelike slices at times $t_+ >
t_-$ (perhaps with $t\pm\to\pm\infty$): one may define initial and
final states of quantum fields in the Heisenberg picture by writing
a wavefunction of field values on $\Sigma_{-}$ and $\Sigma_+$,
respectively. The patch of $AdS$ covered by Poincar\'e coordinates
is the shaded region bounded by ${\cal I}^+\cup{\cal I}^-\cup {\bf
i}^+\cup {\bf i}^-\cup {\bf
i}^0$.}{\epsfxsize1.5in\epsfbox{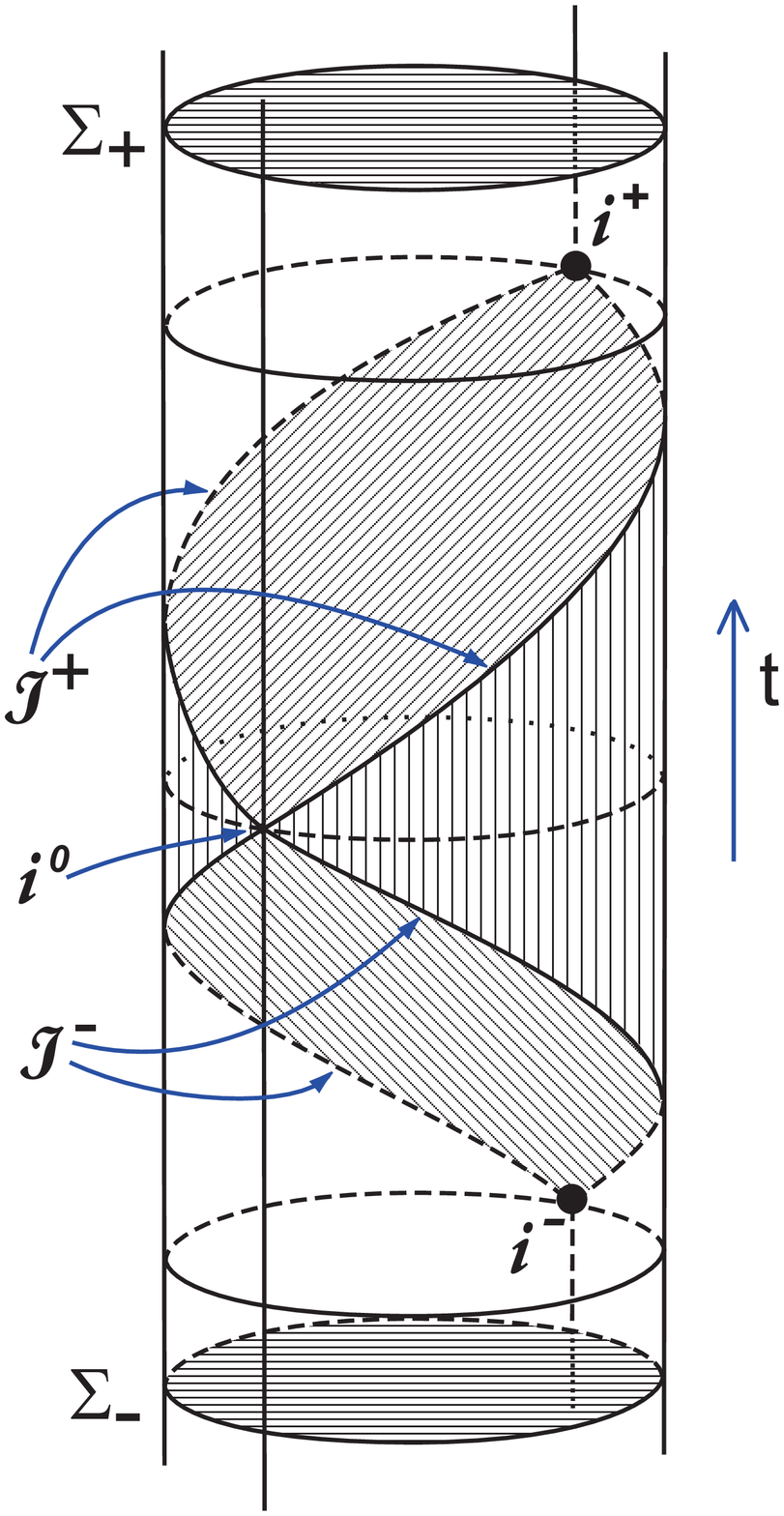}}

Similarly, we can consider the correspondence for the Poincar\'e
patch of $AdS$ space, described by the metric
\eqn\poincoordinates{
    ds^2 = R_{AdS}^2\frac{dz^2 + d{\bf x}^2}{z^2}~,}
where $d{\bf x}^2$ is the metric on four-dimensional Minkowski space
$\IR^{3,1}$. In these coordinates the timelike boundary is at $z =
0$ (see Fig 2.) String theory on this space is dual to the above 4d
CFT on $\IR^{3,1}$.

The natural scales in these compactifications are the string scale
$\ell_s$, the five-dimensional Planck scale $\ell_p$, and the radius
of curvature of the spacetime $R_{AdS}$.  From these we can form two
independent dimensionless ratios, which are dual to dimensionless
parameters in the Yang-Mills theory: $(R_{AdS}/\ell_p)^3 = N^2$ and
$(R_{AdS}/\ell_s)^4 = \lambda \equiv g^2_{YM} N$, where $g^2_{YM}$
is the dimensionless coupling of the gauge theory.

In the large $N$ limit, at fixed Yang-Mills coupling $g_{YM}$, the
low energy supergravity limit of string theory on $AdS_5\times S^5$
is a good approximation. In this situation, local, low-dimension
single-trace operators $\CO_a$ are dual to supergravity fields
$\phi^a$. Among these operators are the 4d stress tensor, which is
dual to the 5d metric in an appropriate gauge.
We will focus on perturbations by these operators, in particular
operators dual to the 5d metric and 5d scalar fields.
Single-trace local operators with dimension of order $\lambda^{1/4}$
are dual to massive string states. Deformations by local multi-trace operators
\refs{\AharonyPA\WittenUA\BerkoozUG-\SeverFK}\ can be described by a
particular deformation of the boundary conditions. The dual
description of Wilson line operators have also been constructed
\refs{\ReyIK,\MaldacenaIM}.

In standard treatments, which we will follow here, scalar fields in
the $5d$ gravitational theory are taken to be dimensionless, and the
bulk effective action scales as $N^2 = (R_{AdS}/\ell_p)^3$. The mass
$m^2$ of these scalars is related to the conformal dimension
$\Delta_{a,+}$ of the dual operators by \refs{\GubserBC,\WittenQJ}:
\eqn\masstodim{
    \Delta_{a,\pm} = 2\pm\sqrt{4 + R_{AdS}^2m^2}~.}
The bound $R_{AdS}^2m^2\ge-4$, required for the operator in the CFT
to have real conformal dimension, coincides with the lower bound on
scalar masses required for stability of the bulk theory
\refs{\BreitenlohnerBM,\BreitenlohnerJF}.

\ifig\angles{The Penrose diagram for $AdS_d$ in Poincar\'e
coordinates.  The boundary $\cal I$ is the timelike boundary of
$AdS$, and is conformal to $\IR^{d-1,1}$. $H_{\pm}$ are coordinate
horizons.  One may specify the quantum states as $t\to \pm\infty$ in
the bulk with data on $\Sigma_\pm=i_{\pm} \cup H_{\pm}$.}
{\epsfxsize1.5in\epsfbox{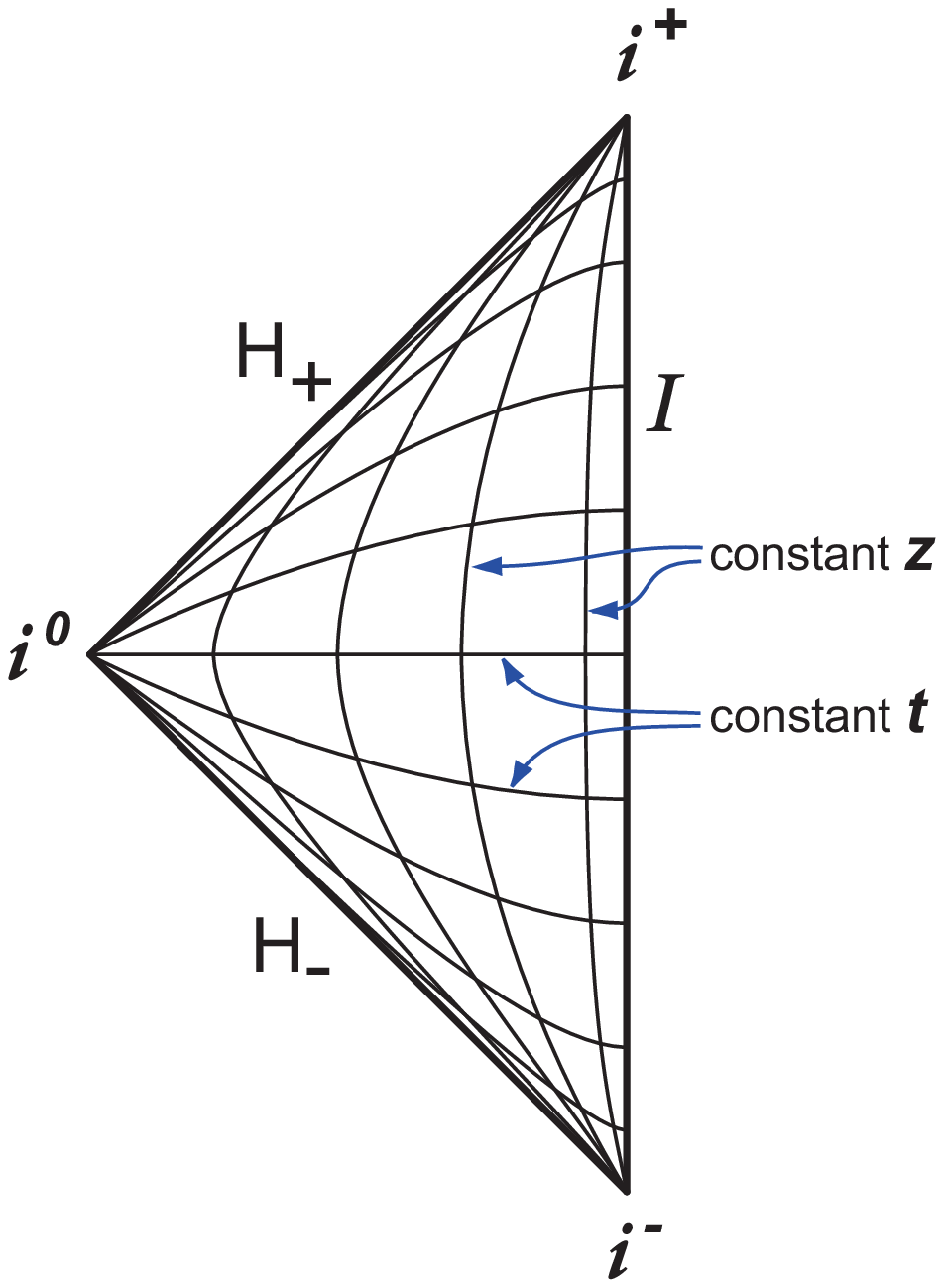}}

\subsec{Boundary behavior of classical fields in AdS}

Consider small fluctuations of classical supergravity scalar fields,
which are well-described by linearized classical supergravity as
$N,\lambda \to\infty$. In Lorentzian signature, at fixed $AdS$
momentum (along $S^3\times \IR_t$ in \globalcoords, or along
$\IR^{1,3}_{{\bf x}}$ in \poincoordinates) there are two independent
solutions to the linearized equations of motion, which are
classified by their boundary behavior as $z\to 0$/$r\to\infty$:
\eqn\boundarybeh{\eqalign{
    \phi^a_1({\bf x},z) & \sim z^{\Delta_{a,-}} \lambda^a({\bf x}) + \ldots
        \ \ ({\rm Poincare})\ \ \ \ \ \phi^a_1(t,\Omega, r) \sim r^{-\Delta_{a,-}}\lambda^a(t,\Omega) \ \ ({\rm global})\cr
    \phi^a_2({\bf x},z)& \sim z^{\Delta_{a,+}} \phi^a_0({\bf x})+\ldots
        \ \ ({\rm Poincare})\ \ \ \ \ \phi^a_2(t,\Omega,r) \sim r^{-\Delta_{a,+}}\phi^a_0(t,\Omega) \ \ ({\rm global})
    ~.}}
A general solution to the linearized equations can be written (using
Poincar\'e coordinates for definiteness) as:
\eqn\gensolone{
    \phi^a({\bf x},z) = \phi^a_1 + \phi^a_2\ .
}
The solution $\phi^a_2$ is normalizable with respect to the standard
Klein-Gordon norm in AdS spacetimes
\refs{\BalasubramanianDE,\BreitenlohnerBM,\BreitenlohnerJF}. The
solution $\phi^a_1$ is normalizable for $-4\le R_{AdS}^2m^2<-3$,
while for $R_{AdS}^2m^2\ge -3$ it is not. As we describe below, the
normalizable modes are candidates for propagating modes in AdS,
while the non-normalizable modes (and normalizable modes when
$R_{AdS}^2m^2 < -3$) are candidate duals to perturbations of the
Hamiltonian of the quantum system.

Higher-spin fields, such as the metric, will be dual to higher-spin operators
on the boundary, such as the stress tensor.
For these modes, a similar story about the boundary behavior
applies.  In the nonlinear supergravity theory, backreaction will couple metric
modes to the scalar modes.  If $m^2 < 0$ for the scalar masses, so that the
dual operators are relevant, the backreaction will be such that the
metric remains asymptotically anti-de Sitter, reflecting the conformal fixed point in the UV.

\subsec{Quantum states in the CFT}

Let us first consider the dynamics of the unperturbed CFT and its
gravitational dual. Assume that we are working at low energies and
at large $N,\lambda$ in the dual field theory, so that the
supergravity approximation in spacetime is valid.

Quantum states of the bulk supergravity at fixed time $t$ can be
represented via wavefunctionals of $\phi^a$:
\eqn\wavefunctionals{
    \Psi\left[\phi^a(t)\right]~,}
In the absence of boundary sources the states must be supported on
field configurations that have boundary behavior specified by the
second line in \boundarybeh\ (we will discuss the ambiguity for
$(mR_{AdS})^2 \leq -3$ at the end of this section.) This can be seen
for small fluctuations of the supergravity fields by building up the
states via second quantization.

Because the AdS/CFT correspondence has a Hamiltonian version
\refs{\WittenQJ}, the Hilbert spaces of the gauge theory and the
dual string theory must be the same.  Furthermore, one can define a
Hamiltonian which has both a gauge theory and a string theory
interpretation. The vacuum can be defined as the state preserving
the $SO(4,2)$ symmetry of the theory (the conformal group of the CFT
or the isometry group of AdS.) The duality map for small
fluctuations of the supergravity fields can be constructed
explicitly by providing a map between the Fourier modes of the CFT
operators $\CO^a$ and the creation and annihilation operators of the
dual supergravity fields $\phi^a$, as in refs.
\refs{\BalasubramanianDE,\WittenQJ,\HorowitzBJ\WittenZW-\BanksDD}.
\foot{In the interacting, finite-$N$ theory, this map must be
modified, as pointed out in \refs{\BanksDD}.  To our knowledge, the
issues raised in that work have yet to be addressed.}

For states defined at past and future infinity, as in \transamp, we
can push the fixed-t slices back to the far past and future, by
defining spacelike hypersurfaces $\Sigma_{\pm}$ at constant global
or Poincar\'e time $t_{\pm}$, as shown in Figs. 1 and 2, and sending
$t_{\pm} \to \pm\infty$. Note that in Poincar\'e coordinates, the
$t_{\pm}\to\pm\infty$ limits of constant-time hypersurfaces are the
unions of the horizons $H_{\pm}$ and timelike infinity ${\bf
i}^{\pm}$.

We are particularly concerned with the case that the supergravity
states are semiclassical coherent states in the bulk,\foot{We use
the term "coherent state" in the sense described by Yaffe
\refs{\YaffeVF}. In \refs{\AharonyTI}\ it is stated that the large-N
limit is not a classical limit in the usual sense, as it corresponds
to a limit with a large number of fields.  However, Ref.
\refs{\YaffeVF}\ gives a precise definition of a classical limit,
and gives a convincing if not complete set of arguments that the
large N limit of a gauge theory is such a classical limit.  The
arguments are independent of the 't~Hooft coupling, and matches our
expectation that this limit is dual to the limit of classical string
theory in anti-de Sitter space. However, we believe that the
arguments in \refs{\YaffeVF}\ require that one take $N\to\infty$
with fixed 't~Hooft coupling $\lambda$. The limit in which
$g^2_{YM}$ is fixed  and small as $N\to \infty$ should be a
different limit, dual to quantum string theory in ten-dimensional
flat space.}
described by macroscopic expectation values for the field operators
$\phi^a({\bf x},z)$.  At leading order in $1/N$, this expectation
value satisfies the supergravity equations of motion and has the
$z\to 0$ behavior of the normalizable modes. At fixed time $t$, such
semiclassical coherent states are well-specified at this order in
$1/N$ by a state in classical phase space: that is, by the value of
the field $\phi^a = \phi^a_0$ and the field momentum $\pi_a \sim
\dot{\phi}_a = \pi_{a,0}$ at fixed time $t$. The quantum
wavefunctional
\eqn\coherentwavefunctional{
    \Psi_{\phi^a_0,\pi_{a,0}}[\phi^a, t] \equiv \brket{\phi^a}{(\phi^a_0,\pi_{a,0})}
}
where $\ket{\phi^a}$ is an eigenstate of the field operator, is
peaked at these values, with a width in phase space proportional to
$\hbar/N^2$.

Consider a solution $\phi^a(x,z,t)$ to the classical equations of
motion with initial conditions $\phi^a_- \equiv \phi^a(t_-)$,
$\pi_{a,-}\equiv \pi_a(t_-)$. If we fix the quantum state
$\ket{\psi_-}$ at time $t_-$ to be a coherent state
$\ket{(\phi^a_-,\pi_{a,-})}$ with the wavefunctionals of $\phi^a$
and $\pi_a$ peaked on these initial conditions at time $t$, then to
leading order in $1/N$ the system will evolve in time through
classical states peaked on $\phi^a(t),\pi_a(t)$.  Now, let
$\ket{\psi_+(t_+)} = \ket{(\phi^a_+,\pi_{a,+})}$.  The transition
amplitude
\eqn\unperttransamp{
    A = \brket{\psi_+(t_+)}{\psi_-(t_-)}~,
}
will be negligible at leading order in $1/N$ unless
$(\phi^a_+,\pi_{a,+}) \sim (\phi^a(t_+),\pi_a(t_+))$ up to
corrections of order $1/N$. Otherwise, there is no semiclassical
trajectory contributing to $A$.

On the other hand, if the initial and final states are eigenstates
$\ket{\phi^a}$ of the field operator, then the transition amplitude
\eqn\unpertpostrans{
    A_{pos} = \brket{\phi^a_+(t_+)}{\phi^a_-(t_-)}~,
}
will generically receive contributions from semiclassical paths
contributing to it for a range of $\phi^a_+,\phi^a_-$. These will be
paths corresponding to solutions to the classical equations of
motion, specified by the initial and final field values
$\phi^a(t_{\pm}) = \phi^a_{\pm}$, as can be deduced from a
stationary phase approximation.  To make contact with the coherent
state approach, one can use the fact that the coherent states form
an overcomplete basis and may be used to construct a resolution of
the identity \refs{\YaffeVF}:
\eqn\coherentresolution{
    {\bf 1} = \int D\phi^a D\pi_a
    \ket{(\phi^a,\pi_a)(t)}\bra{(\phi^a,\pi_a)(t)}~.
}
Let $\phi^a_n(t),\pi_{a,n}(t)$ be the classical positions and
momenta at time $t\in[t_-,t_+]$ consistent with the initial and
final conditions $\phi^a(t_{\pm}) = \phi^a_{\pm}$. The label $n$
takes care of the cases where there may be more than one solution.
If we insert \coherentresolution\ at time $t\in [t_-,t_+]$, we will
find that at leading order in $1/N$, the dominant contributions will
come from the coherent states specified by
$\phi^a_n(t),\pi_{a,n}(t)$.

In order to make contact with the work of
\refs{\BalasubramanianDE,\BalasubramanianSN}, consider the case in
\transamp\ for which $\ket{\psi_\pm}$ are classical coherent
states, consistent with a single classical trajectory $\phi^a(x,t)$.
Then the matrix element of the operator in the dual CFT is specified
by the boundary behavior of the expectation value of the bulk
supergravity field:\foot{For a certain class of operators this
formula will receive corrections, as shown in \refs{\SkenderisUY}.}
\eqn\operatorexp{
    \bra{\psi_+} \CO^a({\bf x}) \ket{\psi_-} = \Delta_+ \phi_0({\bf
    x})~,
}
(Note that if $z$ has length dimension $1$, \operatorexp\ is
dimensionally correct.) To leading order in the $1/N$ expansion, the
classical "coherent" states are completely specified by the
expectation values of classical operators \refs{\YaffeVF}.  The
classical operators are the single-trace operators of the theory
(they may be nonlocal in general.) At this order in $1/N$, every
such operator is independent. For local single-trace operators, one
must specify the expectation value for every frequency and spatial
momentum. Alternatively, one may specify $\phi_0(x)$ in
\operatorexp\ for all $x$.  In the dual theory, with in the linear
approximation to the supergravity equations of motion, this
specifies the classical solution completely. Note that general
(highly quantum) states are not well-characterized by the one point
functions, as discussed for example in~\refs{\PolchinskiYD}.

To be more precise, we must consider the 5d metric coupled to the scalars.
Therefore the classical coherent states are
$\ket{\phi^a,\pi_a;g^{\mu\nu},\pi_{\mu\nu}}$, and the eigenstates of
the field operators are $\ket{\phi^a,g^{\mu\nu}}$.
The topology and geometry of $\Sigma_\pm$ is defined
by the state. This topology can be rather different from a spacelike slice
of AdS spacetime: for example, the spacetime may be an AdS-Schwarzchild
black hole.

\subsec{SUGRA field eigenstates and gravitational backreaction}

The astute reader will worry about our use of eigenstates of the
supergravity field operators. Such states have overlap with
eigenstates of the Hamiltonian states of arbitrarily high energy,
even for a mode of finite frequency. The point is that if $\pi_n$ is
the conjugate momentum for $\phi_n$, the noninteracting Hamiltonian
for this theory is:
\eqn\modehamilt{
   H = \half \pi_n^2 + \omega_n^2 \phi_n^2 + ...
}
If the uncertainly in $\phi_n$ vanishes, the uncertainty in $\pi_n$
is infinite, and so the energy uncertainty is infinite.  This is
potentially disastrous when coupling the theory to gravity.

Nonetheless, we believe that we are safe so long as we calculate
objects such as \unpertpostrans.  First, in the dual CFT there is no
gravity, and so there seems to be no problem in principle in
considering states of as high an energy as we please. In the
supergravity theory, a state of fixed but large energy, made up of
normalizable modes, will not change the asymptotic structure of
anti-de Sitter space.  A black hole is the generic example of such a
state, and black hole solutions do not disturb the asymptotic
structure at timelike infinity.

Secondly, from the spacetime point of view, $\ket{\phi}$ is not a
classical coherent state, and it makes no sense to simply insert
$\bra{\phi}T_{\mu\nu}\ket{\phi}$ into Einstein's equations in order
to compute the backreaction.  Instead, one can decompose this state
via \coherentresolution:
\eqn\positionresolution{
   \ket{\phi} = \int D\phi_0 D\pi_{0} \ket{(\phi_0,\pi_{0})}
   \brket{(\phi_0,\pi_{0})}{\phi} \equiv \int D\phi_o D\pi_0
    \ket{(\phi_0,\pi_{0})} \Psi_{(\phi_0,\pi_0)}(\phi)\ .
}
Here $\Psi_{\phi_0,\pi_0}(\phi)$ is the wavefunctional for a
classical coherent state with the expectation values of $\phi,\pi$
peaked on $\phi_0,\pi_0$, and the peak has width $1/N$.  The energy
of such a state is also sharply peaked at its classical value.  In
the classical limit, one should compute the classical gravitational
backreaction for each such classical state.  In other words,
\coherentresolution\ should really be considered as an integral over
classical states of the scalars and the metric. For each such
classical state, time evolution will generate a classical spacetime.
Some of these spacetimes will be strongly gravitating. Generically
they will be black holes of arbitrarily high energy. But if we
compute \unpertpostrans\ for sufficiently weak fields, these highly
energetic states will not contribute.

\subsec{Perturbing the CFT}

Next, let us consider the case when the $\CN=4$ SYM action is
perturbed by local scalar operators. The case where the action is
perturbed by the stress-tensor (which is dual to the bulk metric)
will not be considered. The basic prescription is stated in
\refs{\GubserBC,\WittenQJ}: a perturbation of the CFT Hamiltonian by
a (spacetime-dependent) coupling $\lambda^a(x)$ is dual to
performing the path integral over supergravity modes $\phi^a({\bf
x}, r)$ with boundary conditions at timelike infinity specified by
the first line of \boundarybeh.  In the classical, large-N limit, a
given classical solution satisfying these boundary conditions will
be a saddle point solution describing a transition amplitude between
two coherent states.


\subsubsec{The limit of linearized supergravity}

Let us first consider scalar fields which remain small in the
interior of the AdS spacetime. In this case bulk interactions can be
neglected and a general solution $\phi^a({\bf x},z)$ with the $z\to
0$ behavior dominated by the first line of \boundarybeh\ can be
written as:
\eqn\classicalfields{ \eqalign{
    \phi^a(x,z) & = \int d^4 x' G^a_{\p B}(x,z;x')\lambda^a(x') + \phi_{v}(z,x) \cr
    \phi_{v}(x,z) & \to_{z\to 0} z^{\Delta_+} \widetilde\phi(x)\ .
}}
Here $G$ is the bulk-boundary propagator in the AdS vacuum, as
defined in \refs{\WittenQJ}\ and more carefully in
\refs{\FreedmanTZ}. Note that even in the limit of linearized
supergravity, the map between and $\widetilde\phi$ and the field
eigenstates at $\Sigma_{\pm}$ will depend on the coupling: this is
because the first term on the right hand side of \classicalfields\
has support on $\Sigma_{\pm}$. Therefore, to keep the states at
$\Sigma_{\pm}$ fixed while changing $\lambda$, we must also change
$\widetilde\phi$. On the other hand, changing the state will change
$\widetilde\phi$ and not $\lambda$.

The normalizable piece which scales as $\phi_2$ in \boundarybeh\ is
a linear combination of $\lambda$ and $\widetilde\phi$.  In
Poincar\'e coordinates, any linear combination is allowed when the
momenta dual to ${\bf x}$ is timelike, corresponding to the
existence of propagating states for any such momentum. For spacelike
momenta there are not propagating states, and the normalizable and
non-normalizable modes must come in a specific linear combination in
order to avoid a singularity in the interior of AdS
\refs{\GubserBC,\WittenQJ}.

We should note that many singularities are resolvable in that they
reflect interesting infrared physics in the dual quantum field
theory, or D-brane sources in the interior of spacetime. However, it
is important that not all such singularities are resolvable
\PolchinskiUF. We leave this question, in the context of our
discussion of holographic renormalization, for future work.

In the end, in addition to the piece of $\phi_2$ which depends on
$\lambda$, we may add a piece that is specified by $\widetilde\phi$
in \classicalfields. (In global coordinates, the considerations
above restrict the frequency decomposition of $\widetilde\phi$). This
freedom has a reflection in the field theory: the one-point function
at finite $\lambda$ in the noninteracting, large-N limit is
\refs{\BalasubramanianDE}:
\eqn\onepointfunction{
    \vev{\CO({\bf x})} = \Delta_+ \widetilde\phi + c \Delta_+ \int d^4{\bf x}'
        \frac{\lambda({\bf x}')}{|{\bf x} - {\bf x}'|^{2\Delta_+}}~,
}
where $c$ is a constant independent of $\lambda,\widetilde\phi$, and
we have chosen to state the results in Poincar\'e coordinates.

\subsubsec{Perturbations in the interacting theory}

We will be interested in relevant perturbations which grow in the
infrared. For perturbations which become large in the IR
the bulk interactions cannot be neglected and the spacetime
can change drastically.  This leads to two issues that we need to address.

First, for \transamp\ to make sense as a
generating functional for correlation functions, the states should
be independent of variations of the couplings $\lambda^a(x)$. Given
our experience with solutions to the linearized equations, we might be tempted to
define the states via the expectation values of operators, which
depends on the piece of the scalar fields behaving as $\phi_2$ in
\boundarybeh, or via $\widetilde\phi$ in \classicalfields.  However,
once we take nonlinearities of classical supergravity into account,
the map between the quantum state and this data will depend on
$\lambda(x)$.

These problems are avoided (in the supergravity approximation) by defining the states via
wavefunctionals of $\phi^a$ (and $g^{\mu\nu}$) on spacelike slices
$\Sigma_{\pm}$ at fixed times $t_{\pm}$ \refs{\MarolfFY}, so long as
one only varies the couplings strictly in the interior of $t_{\pm}$.
Since we have defined a set of states in
the bulk in a way that is independent of the boundary conditions at
timelike infinity, duality implies we have defined a set of states
which are independent of the $4d$ couplings.  However, describing
these states as $4d$ field theory operators acting on the vacuum may
be very difficult.  To begin with, the geometry and topology of $\Sigma_{\pm}$
may be very different from a spacelike slice of AdS.  For example,
upon perturbing the theory by a mass term, an "infrared wall" may develop
at finite radius \refs{\PolchinskiUF}.\foot{We would like to thank O. Aharony
for correspondence on this issue.}  In such a situation, the states must
define a slice $\Sigma_{\pm}$ of the appropriate topology, or there
will be no semiclassical trajectory contributing to \transamp. Note that
near the wall, one must specify more than the values of the supergravity
fields in order to define the state: classical supergravity breaks down near such a wall,
and the singularity is resolved by stringy and quantum effects.

\subsec{The generating function of correlation functions}

Given the prescription above, we can now construct all elements of
\transamp. In the remainder of this paper, we wish to consider
\transamp\ as a solution to the classical Hamilton-Jacobi equations
of the supergravity theory. In this formulation, the constants of
motion will essentially be the couplings and the states. This map
will make sense if, for fixed $\ket{\psi_{\pm}}$ and $\lambda^a$,
there is a unique classical saddle point in the path integral
representation of \transamp\ for every small variation of
$\lambda^a$. As we have stated above, this will not be true if we
choose $\ket{\psi_{\pm}}$ to be definite classical coherent states.
Instead, we will choose $\ket{\psi_{\pm}} =
\ket{\phi^a(t_{\pm}),g^{\mu\nu}(t_\pm)}$ to be eigenstates of the
field operators.  In this case, the generating function of
correlation functions in the large-$N$ limit is the classical action
\eqn\classicalgenerator{
    Z[\{\lambda\}] = \exp\left(\frac{i}{\hbar} S_{cl}[\lambda^a({\bf
    x}),(\phi^a_-,g^{\mu\nu}_-),(\phi_+^a,g^{\mu\nu}_+)]\right)~,
}
where $S_{cl}$ is the classical supergravity action evaluated
between $\Sigma_{\pm}$, on solutions to the classical equations of
motion such that
\eqn\bouncond{ \eqalign{
    \phi({\bf x},r\to\infty) & \sim r^{-\Delta_{-,a}}\lambda^a({\bf x}) \cr
    \phi^a(t_\pm) = \phi^a_{\pm}~,&\quad
    g^{\mu\nu}(t_\pm)=g^{\mu\nu}_\pm~,
}}
and the metric is asymptotic AdS.

A small variation of \classicalgenerator\ with respect to $\lambda$ will lead
to boundary terms at the timelike boundaries only. Since the value
of the bulk fields is fixed at $\Sigma_\pm$, the variation there
vanishes by construction. In \S5\ this point will be important in claiming that $S_{cl}$ in
\classicalgenerator\ solves
the Hamilton-Jacobi equation.  For more general states we can compute
correlation functions by integrating
$\phi_{\pm}$ over some wavefunctionals which are sharply peaked on
states of finite energy.  This leads to the prescription in \refs{\MarolfFY}.
For example, one can suppress the high-energy fluctuations discussed in \S3.4\ by
choosing the states to be described by smooth (\eg\ gaussian) wavefunctionals peaked about
the field eigenstates. In our proposal, one varies the
Hamiltonian, and then one integrates over initial and final states
in order to suppress the high-energy contributions.  We are then assuming
that the integral with respect to $\phi_{\pm}$ and the derivatives
with respect to $\lambda(x)$ commute.

Before closing we must point out an additional subtlety in this
discussion. In the range $-4\le m^2<-3$, both $\phi_{1,2}$ in
\boundarybeh\ are normalizable, and the identification of these with
$\lambda^a$, $\widetilde{\phi}^a$ can be reversed
\refs{\BalasubramanianDE,\KlebanovTB}. The generating functions of
correlation functions of the two theories are related by a Legendre
transformation \refs{\KlebanovTB,\HartmanDY}. In this work we will
take the solution scaling as $z^{\Delta_-}$ to correspond to the
coupling, although it is not always natural to do so (\cf\ \S2\ of
\refs{\BalasubramanianSN}.)

\newsec{Holographic renormalization and the Hamilton-Jacobi equation}

In this section we will embark on a critical review and discussion
of the results of \refs{\deBoerXF,\deBoerCZ}, in order to better
explain and eventually answer the questions raised in the introduction. We
will restate the results of those papers in some detail, as we will
need to comment on some specific points.

\subsec{General discussion}

As pointed out by various authors, beginning with
\refs{\WittenQJ,\FreedmanTZ}, computations of correlators in both
the bulk and boundary theories require regularization. The bulk
calculations contain divergent terms in $S_{SUGRA}$ arising from the
$r\to\infty$/$z\to 0$ region of the spacetime. There is by now a
well-defined procedure for subtracting these divergences in the bulk
and interpreting this subtraction procedure as a choice of local
ultraviolet counterterms in the dual field theory (see for example
\refs{\SkenderisWP}\ for a review and references.) In the field
theory, the counterterms determine the beta functions of the theory.
The result is a supergravity expression for the objects driving the
RG flow in the dual field theory.

This suggests a regularized version of the AdS/CFT correspondence
\refs{\BalasubramanianJD}. Consider the correspondence for the
Poincar\'e patch of $AdS$ spacetime. The classical Lagrangian of the
bulk supergravity theory is integrated over $z > z_{UV}$, to define
\eqn\cutoffact{
    S_{reg}(\phi_{UV},g_{UV}; z_{UV}) = \int_{z \geq z_{UV}} d^5 x
        \CL(\phi, g)~,
}
where the Lagrangian is evaluated on solutions to the classical
equations of motion, with boundary values
\eqn\newbc{
    \phi_{UV}({\bf x}) = \phi({\bf x}, z_{UV})\ ;\ \ \ \ \ g_{UV,\mu\nu}({\bf x}) =
        g_{\mu\nu}({\bf x}, z_{UV})~.
}
As with the "unregulated" version of the correspondence, this
prescription is well-defined in Euclidean space because the
classical equations of motion with boundary conditions \newbc\ have
a unique nonsingular solution for $z > z_{UV}$.\foot{Again,
some additional singular solutions are allowed in Euclidean space. These have definite physical interpretations and introduce no ambiguity in the interpretation of \cutoffact.}
After subtracting a set of
counterterms, $S_{reg}$ is identified with the generating function
of correlators in the dual theory, cut off at an energy scale
proportional to $z_{UV}^{-1}$~\refs{\SusskindDQ} (although it will
correspond to a fairly complicated cutoff prescription
\refs{\BalasubramanianJD}.) In this picture, $\phi_{UV}, g_{UV}$ are
dual to couplings in the cutoff theory. The evolution of the fields
$\phi_{UV}, z_{UV}$ as one increases $z_{UV}$ is expected to be dual
to the renormalization group flow of the boundary couplings, as one
lowers the UV cutoff, and it can be shown that the evolution of this
generating function with $z_{UV}$\ is governed by a kind of
Callan-Symanzik equation~\refs{\BalasubramanianDE}.

The radial evolution of $S_{reg}$, $\phi_{UV}$, and $g_{UV}$ with
$z$ can be described by the Hamilton-Jacobi equation. In the
asymptotic limit $z\to 0$, the radial Hamilton-Jacobi equation can
be rewritten as a set of Callan-Symanzik equations for the boundary
correlators \refs{\deBoerXF,\deBoerCZ}, via a construction we now
review and discuss.

\subsec{The radial Hamilton-Jacobi equation}

In classical general relativity coupled to matter, the Hamiltonian
constraint $\CH=0$ is a Hamilton-Jacobi equation equation of the
form $K({\bf a},{\bf b},t)=0$ \Hamilton. The evolution of the fields
$\phi, g$ can then be computed via either \hjdiffeqs\ or \bconst. We
will use \hjdiffeqs, and find that it has a close relationship to
the RG equations of the dual field theory.

De Boer {\it et. al.}\ consider the Euclidean AdS/CFT
correspondence. The Euclidean metric can be written using an ADM
decomposition:
\eqn\admmetric{
    ds^2 = R_{AdS}^2 \CN^2 dr^2 + R_{AdS}^2 g_{\mu\nu}\left(dx^{\mu} + \CN^{\mu} dr\right)
        \left(dx^{\nu} + \CN^{\nu} dr\right)~,}
where $\CN,\CN^{\mu}$ are the lapse and shift functions. Locally, we
can use the diffeomorphism invariance to choose $\CN=1$ and
$\CN^{\mu}=0$, and we will work in this gauge from now on. (In the
Poincar\'e patch of AdS the metric in this gauge is related to
\poincoordinates\ by $z=R_{AdS}\ e^{-r}$.) However, after gauge
fixing we must still impose the equations of motion for $\CN$ and
$\CN^{\mu}$.  These give rise to the Hamiltonian constraint and the
diffeomorphism constraints, respectively.

Note that we have put an explicit factor of $R_{AdS}^2$ in front.
With this normalization, the factors of $R_{AdS}$ in the classical
supergravity action appear in the combinations $R_{AdS}/\ell_{s} =
\lambda^{1/4}$ and $R_{AdS}/\ell_{p,5} = N^{2/3}$, where $\lambda =
g_{YM}^2 N$ is the 't Hooft coupling.  However, it means that
$g_{\mu\nu}$ has mass dimension 2.

We write Hamilton's principal function
\eqn\fivedhjsolution{
    S[(g_{\mu\nu}({\bf x}),\phi({\bf x})),{\bf a}]\ ,
}
as a functional of the configuration space variables
$(g_{\mu\nu}({\bf x}),\ \phi({\bf x}))$, and the constants of motion
${\bf a}$. In doing so we must take symmetry under diffeomorphisms
into account. Two metrics which differ only by a coordinate
transformation describes the same point in the configuration space.
The diffeomorphism constraint:
\eqn\bulkconstraints{
    \nabla^{\mu} \frac{\delta S}{\delta g^{\mu\nu}} + \frac{\delta S}{\delta \phi^a}
        \nabla_{\nu} \phi^a = 0~,}
ensures that $S$ is invariant under 4-d coordinate transformations
and therefore is a good function on the configuration space.

The Hamiltonian constraint $\CH = 0$ is
\eqn\hamiltoniancon{
    \frac{1}{\sqrt{g}} \left[ \frac{1}{3} \left(
    g^{\mu\nu}\frac{\delta S}{\delta g^{\mu\nu}}
        \right)^2 - g^{\mu\lambda}g^{\nu\rho}
            \frac{\delta S}{\delta g^{\mu\nu}} \frac{\delta S}{\delta g^{\lambda\rho}}
            - \half G^{ab}(\phi) \frac{\delta S}{\delta \phi^a}\frac{\delta S}{\delta \phi^b}
            \right] = \CL~,}
where
\eqn\scalarbulklag{
    \CL = \sqrt{\hat g}\left(
        \half G_{ab}(\phi) \hg^{IJ}\p_I \phi^a \p_J \phi^b +\hat\CR - V(\phi)
        \right)~,}
is taken to be the (bosonic) $5d$ Lagrangian for minimally coupled
scalar fields $\phi^a$ and bulk 5d metric $\hat{g}^{IJ}$
($I,J=1\dots 5$). The constants of motion ${\bf a}$ in
\fivedhjsolution\ parametrize solutions to
\hamiltoniancon,\bulkconstraints; we will discuss them below.
In a gravitational theory, $S$ does not depend explicitly on $r$.  Eq. \hamiltoniancon\ is
the Hamilton-Jacobi equation in this context.  Once
we have solved for \hamiltoniancon\ for $S$, the equations of motion for $g$ and $\phi$
following from Hamilton's equations are:
\eqn\hjbulkeom{ \eqalign{
    \frac{\p\phi^a({\bf x},r)}{\p r} & = \frac{G^{ab}(\phi)}{\sqrt{g}}\frac{\delta S}{\delta\phi^b({\bf x},r)}\cr
    \frac{\p{g}_{\mu\nu}({\bf x},r)}{\p r} & = \frac{1}{\sqrt{g}}\left(- 2 \frac{\delta S}{\delta{g}^{\mu\nu}({\bf x},r)}
        + \frac{2}{3} g_{\mu\nu}g^{\lambda\rho} \frac{\delta S}{\delta{g}^{\lambda\rho}({\bf x},r)}
        \right)
~.}}

\subsec{Solving the Hamilton-Jacobi equations}

Let us consider the case with only marginal and relevant
perturbations of the CFT, following \refs{\deBoerXF,\deBoerCZ}. We
will also follow these references and consider solutions $S$ in the
region $r\to\infty$, dual to the UV regime of the field theory. In
that limit the authors of \refs{\deBoerXF,\deBoerCZ} propose the
following path to a solution. They write $S$ and $\CL$ in a
derivative expansion
 \eqn\devexp{S=S^{(0)}+S^{(2)}+\Gamma\ ,\qquad \CL=\CL^{(0)}+\CL^{(2)}~,}
where
 \eqn\actionbreakup{\eqalign{
    S^{(0)} &= \int d^4x \sqrt{g}\ U(\phi)\cr
    S^{(2)} & = \int d^4x\sqrt{g} \(
        \Phi(\phi)\CR + \half M_{ab}(\phi)g^{\mu\nu}
            \p_{\mu}\phi^a \p_{\nu}\phi^b\)\cr
    \CL^{(0)}&=-\sqrt{\hat g}\ V\cr
    \CL^{(2)}&=\sqrt{\hat g}\(\hat\CR + \half G_{ab}\hat g^{IJ}\p_I\phi^a\p_J\phi^b\)~.}}
Here $\CR$ is the $4d$ Ricci curvature for $g$ and ($U$, $M_{ab}$,
$\Phi$) are some functions of ($\phi({\bf x}),g_{\mu\nu}({\bf x})$)
to be determined from \hamiltoniancon. $\Gamma$ is the nonlocal part
of $S$. Note that in this discussion we are assuming that the
spacetime effective action $\CL$ is captured by the zero- and
two-derivative terms. This is dual to the statement that we work to
leading order in $1/N$ and $\lambda^{-1/4}$.

Next, one solves \hamiltoniancon\ order by order in the derivative
expansion.  If we define
\eqn\poisson{
    \left\{A,B\right\}\equiv\frac{1}{\sqrt{g}}\left[\frac{1}{3}\left(
        g^{\mu\nu}\frac{\delta A}{\delta g^{\mu\nu}}\right)
        \left(g^{\lambda\rho}\frac{\delta B}{\delta g^{\lambda\rho}}\right)
            - g^{\mu\lambda}g^{\nu\rho}
            \frac{\delta A}{\delta g^{\mu\nu}}\frac{\delta B}{\delta g^{\lambda\rho}}
            - \half G^{ab}\frac{\delta A}{\delta\phi^a}\frac{\delta B}{\delta
            \phi^b}\right]~,}
then the Hamilton-Jacobi equation breaks up into a set of equations
for each order in the derivative expansion.  These are written in
\refs{\deBoerXF,\deBoerCZ}\ as:
\eqn\gradedsolutions{ \eqalign{
    & \left\{S^{(0)},S^{(0)}\right\}  = \CL^{(0)} \cr
    & 2 \left\{S^{(0)}, S^{(2)}\right\}  = \CL^{(2)} \cr
    & 2 \left\{S^{(0)},\Gamma\right\} + \left\{S^{(2)},S^{(2)}\right\} = 0~.}}
This scheme for solving the Hamilton-Jacobi equations via a
derivative expansion is sensible as $r\to\infty$ because the various
terms in \devexp\ scale differently in this limit: terms with lower
number of derivatives diverge more rapidly. The result is a clean
interpretation of $S$. $S^{(0)}$, $S^{(2)}$ correspond to divergent
counterterms; a study of explicit solutions to \gradedsolutions\
shows that they are local. $\Gamma$ is the spacetime effective
action minus these counterterms, that is, the regularized generating
function of correlators in the dual field theory.\foot{The second
term in the last line of \gradedsolutions\ corresponds to the
gravitational anomaly in four dimensions, and the first term
includes the expectation value of the trace of the stress tensor.
Therefore it makes sense to assign $\Gamma$ dimension 4 in the
derivative expansion. For a generalization to other dimensions see
\KalkkinenVG.}

This procedure for solving \hamiltoniancon\ is adapted to the limit
$r\to\infty$.  Furthermore, following \refs{\deBoerXF,\deBoerCZ}, we
have ignored the terms $\left\{S^{(2)}, \Gamma\right\}$ and
$\left\{\Gamma,\Gamma\right\}$\  that appear in the full
Hamilton-Jacobi equations, as they are negligible in the
$r\to\infty$ limit. An extension of the formalism to finite $r$ is
desirable if one wishes to study trajectories of renormalization
group flows over a range of scales. We will return to this issue in
\S5.5.

Even in this limit, there is an additional interpretational problem
with this method for solving the Hamilton-Jacobi equations. As we
discussed in \S2, there are many possible solutions to the equations
of motion, corresponding to different choices of constants of motion
{\bf a}. On the other hand, if we desire solutions which are
nonsingular in the interior $r < \infty$ of Euclidean AdS
spacetimes, no such freedom exists \refs{\WittenQJ}.  The constants
of motion ${\bf a}$ represent the states in the Lorentzian
correspondence. The QFT dual of this will become apparent in \S5.
Correlation functions in Euclidean space rotate to vacuum
correlators: in other words, a particular state is selected.

The classical action \cutoffact\ will solve the Hamilton-Jacobi
equation and generate solutions which are nonsingular in the
interior.  For the formalism of \refs{\deBoerXF,\deBoerCZ}\ to match
the results of \refs{\GubserBC,\WittenQJ}, it must be true that as
$r\to\infty$ ($z_{UV}\to 0$), \cutoffact\ can be written as \devexp\
and is a solution to \gradedsolutions.

All of this said, other solutions to the Hamilton-Jacobi equations
exist, even in the case of Euclidean signature.  The solution to
these equations will not have an interpretation as the generating
function of correlation functions.  We will discuss one such
solution, and its interpretation, in \S5.4.

\subsec{Relation to the renormalization group}

\subsubsec{Beta functions}

Near the boundary $r\to\infty$ the bulk fields behave as $\phi^a
\sim e^{-\Delta_- r}$ and are dual to the coupling constants of the
field theory. The evolution of $\phi$ with $r$ should be related to
the running of the dual coupling under the renormalization group.
The bulk fields are in general spacetime dependent, so we should
work with spacetime-dependent couplings
\refs{\OsbornGM,\ErdmengerJA}. However, if the couplings are slowly
varying in ${\bf x}$, then they can be treated as constant in the
UV, and we should recover the standard RG equations in that limit.
The bulk dual of this statement is that as $r\to\infty$, $S^{(0)}$
dominates over the higher-derivative and nonlocal terms in $S$.
Therefore, let us first consider solutions which are constant in
${\bf x}$.

The first equation in \gradedsolutions\ is
 \eqn\UV{V=\frac13 U^2-\half G^{ab}\d_aU\d_bU~,}
and determines $U$ in terms of $V$ (up to terms of total dimension
4). Eq. \hjbulkeom\ then becomes
\eqn\constantflow{ \eqalign{
    \p_r \phi^a & = G^{ab}\p_b U \cr
    \p_r g_{\mu\nu} & = - \frac{1}{3} U(\phi)g_{\mu\nu}
~,}}
where both $\d_r\phi$ and $\d_r g$ have corrections which can be
neglected near the boundary. The second equation in \constantflow\
can be solved with the ansatz $g_{\mu\nu}(r,\bx) =
\rho^2(r,\bx)\tilde g_{\mu\nu}(\bx)$. Here $\tilde g$ is
dimensionless and independent of $r$, while the scale factor $\rho$
has mass dimension $1$ and satisfies
 \eqn\prefact{\d_r\ln(R_{AdS} \rho)=-\frac16 U(\phi)~.}
The study of bulk probes in the AdS/CFT correspondence shows that
the rescaling in the boundary theory is simply related to the
rescaling of $g$ in the bulk. Furthermore, since the solutions
$\phi_1$ in \boundarybeh\ dominate as $r\to\infty$, we can identify
the coupling $\lambda^a$ at some UV scale $\rho$ with the dual field
$\phi(r(\rho))$, up to a power of $\rho$.\foot{If $m^2 R_{AdS}^2
\leq -3$ then this is not always true, as noted in \S2.2.  However,
in Euclidean space, $\phi_1$ and $\phi_2$ are proportional in
Fourier space, and related by convolution with the boundary Green
function in position space (\cf\ \refs{\KlebanovTB}.)} Following
these two observations, de Boer {\it et. al.}\ propose that the beta
function be defined as:
\eqn\phiunderscalechange{
    \beta^a\equiv \rho \p_{\rho} \phi^a = \frac{1}{\p_r \ln (\rho)} \p_r \phi^a
    = - \frac{6}{U(\phi)} G^{ab} \p_b U(\phi)~.
}
Because $\phi \sim z^{\Delta_{-,a}} \lambda_a(\bx)$ is
dimensionless, we identify $\beta^a$ as the beta functions for the
dimensionless coupling. (The difference between this and the beta
function for $\lambda$ is that the latter will have no linear term.)

Let us study the solutions to these equations in more detail. The
spacetime effective potential has the form
\eqn\stpotential{
    V = 12 - \half m_a^2 (\phi^a)^2 + g_{abc}\phi^a \phi^b \phi^c\ .
}
If $G_{ab} = \eta_{ab} + O(\phi^2)$ then the solution for $U$ is
given by \refs{\deBoerXF,\deBoerCZ}:
 \eqn\ufunc{U=-6-\half\vartheta_a\phi^a\phi^a+\frac{g^a_{bc}}{8 - \Delta_a - \Delta_b -
 \Delta_c}\phi^a\phi^b\phi^c~.}
(Note that in \refs{\deBoerXF,\deBoerCZ} the sign in front of the
first term of the right hand side of \ufunc\ is opposite to that in
\stpotential. We have checked that the signs here are self
consistent.) Here $\vartheta_a$ is related to $m_a^2$ by
 \eqn\num{\vartheta_a^2-4\vartheta_a=m_a^2~.}
Choosing the root $\vartheta_a=4-\Delta_a$, leads to the beta
function:\foot{In \S5.4\ we will discuss the other root of \num.}
\eqn\betaquadratic{
    \beta^a = - (4 - \Delta_a) \phi^a -\frac{g^a_{bc}}{8 - \Delta_a - \Delta_b - \Delta_c}
        \phi^b \phi^c~.
}
If the OPE coefficients
\eqn\opealg{
    \CO_b(x)\CO_c(y) \sim \frac{C^a_{bc}}{|x-y|^{\Delta_b + \Delta_c - \Delta_a}}
        \CO_a(y)~,
}
are equal to
\eqn\opeform{
    C_{bc}^a = - \frac{2 g^a_{bc}}{S_3 \left(8 - \Delta_a - \Delta_b -
    \Delta_c\right)}~,
}
where $S_3$ is the volume of a unit 3-sphere,
then \betaquadratic\ is precisely of the form derived, for constant
couplings, in for example \refs{\CardyXT}. This is an unsurprising
answer, as we expect the three-point functions in the bulk to be
related to the boundary OPEs.

The actual relationship between $g_{abc}$ and the boundary OPE
coefficients was calculated in \refs{\FreedmanTZ}.  They differ by a
complicated ratio of gamma functions.  However, unless the OPEs in
question are "resonant", such that $\Delta_b + \Delta_c - \Delta_a =
4$, the quadratic term in the beta functions are scheme dependent.
In fact, standard RG schemes such as minimal subtraction will lead
to quadratic terms in the beta function only if there are associated
divergences, which happens when $\Delta_b + \Delta_k - \Delta_a \geq
4$ in \opealg.\foot{See for example \refs{\FreedmanWX}\ for an
extensive discussion of the scheme dependence of conformal
perturbation theory in two dimensions.  Most of the basic lessons
lift to higher dimensions.} At best we only expect universal answers
at quadratic order in $\lambda$ and to zeroth order in $4 - \Delta_a
= \epsilon_a \ll 1$. Our conclusion is that the holographic RG
calculation outlined in \refs{\deBoerXF,\deBoerCZ}\ correspond to a
particular choice of scheme that is closer to the schemes used in
refs. \refs{\CardyXT,\ZamolodchikovTI}.  These schemes are natural
and useful in conformal perturbation theory; in particular they are
useful for studying the approach to nontrivial infrared fixed
points.

\subsubsec{The Callan-Symanzik equation}

According to the AdS/CFT correspondence \transamp,
 \eqn\gener{\Gamma=\ln\(Z\)}
is the generating function of connected correlation functions, in
the limit $r\to\infty$ or $z\to 0$, if we assume that $S =
S_{SUGRA}$. According to de Boer \etal, the asymptotic correlation
functions are:
 \eqn\correl{
 \<O_{a_1}({\bf x}_1)O_{a_2}({\bf x}_2)\dots O_{a_n}({\bf x}_n)\>_c
    { = \atop ?} {1 \over\sqrt{g({\bf x}_1)}}{\delta\over\delta\phi^{a_1}({\bf
 x}_1)}\dots{1 \over\sqrt{g({\bf x}_n)}}{\delta\over\delta\phi^{a_n}({\bf
 x}_n)}\ \Gamma~,}
where $\<\dots\>_c$ is the connected piece of the correlator.
However, with our normalization, $\phi$ is dimensionless and so
$\delta/\delta\phi(x)$ has mass dimension $d$.  $g_{\mu\nu}$ has
mass dimension $2$, so that the right hand side above is
dimensionless.  If we wish $\CO_{a_k}$ to correspond to operators of
dimension $\Delta_{+,a_k}$, we need to modify this expression. We
conjecture that the correct asymptotic expression is:
\eqn\realcorrel{ \eqalign{
    \<O_{a_1}({\bf x}_1)O_{a_2}({\bf x}_2)\dots O_{a_n}({\bf x}_n)\>_c
    & = {\rho^{\Delta_{+,a_1}} \over\sqrt{g({\bf x}_1)}}{\delta\over\delta\phi^{a_1}({\bf
 x}_1)}\dots{\rho^{\Delta_{+,a_n}} \over\sqrt{g({\bf x}_n)}}{\delta\over\delta\phi^{a_n}({\bf
 x}_n)}\ \Gamma\cr
    & = \frac{\rho^{-\Delta_{-,a_1}}}{\sqrt{\tilde g}({\bf x}_1)}
    \frac{\delta}{\delta\phi^{a_1}({\bf x}_1)}
    \ldots \frac{\rho^{-\Delta_{-,a_n}}}{\sqrt{\tilde g}({\bf x}_n)}
    \frac{\delta}{\delta\phi^{a_n}({\bf x}_n)}\ \Gamma\ .
}}
As evidence, we will find below that the Hamilton-Jacobi equations
imply the statement that the left hand side of \realcorrel\
satisfies the Callan-Symanzik equation. Furthermore, $\rho\to 1/z$
as $z\to 0$, and $\phi^a \to z^{\Delta_{-,a}}\lambda^a$. Therefore,
using \boundarybeh, we can rewrite \realcorrel\ as:
\eqn\othercorrel{
    \<O_{a_1}({\bf x}_1)O_{a_2}({\bf x}_2)\dots O_{a_n}({\bf x}_n)\>_c
    = \frac{1}{\sqrt{{\tilde g}({\bf x}_1)}}{\delta\over\delta\lambda^{a_1}({\bf
 x}_1)}\dots\frac{1}{\sqrt{{\tilde g}({\bf x}_n)}}{\delta\over\delta\lambda^{a_n}({\bf
 x}_n)}\ \Gamma~,}
which is the expected definition of the correlation
functions.\foot{Eq. \othercorrel\ also follows from rewriting a
general conformal correlation function in a diffeomorphism invariant
way using the metric $g_{\mu\nu}$ instead of $\tilde g_{\mu\nu}$
(which corresponds to \correl) and then plugging the relation
$g_{\mu\nu} = \rho^2(r)\tilde g_{\mu\nu}$.}

The third equation in \gradedsolutions\ essentially is a local form
of the Callan-Symanzik equation \refs{\OsbornGM,\ErdmengerJA}. Using
the above result, it becomes:
\eqn\cslikeequation{
    {1\over Z}\[\rho({\bf x}) \frac{\delta}{\delta \rho({\bf x})}
        + \beta^a({\bf x}) \frac{\delta}{\delta \phi^a({\bf x})}\] Z = - \frac{6\sqrt{g}}{U\(\phi({\bf x})\)}
        \left\{S^{(2)}({\bf x}),S^{(2)}({\bf x})\right\}~.
}
The right hand side is the contribution of the conformal anomaly
\refs{\KalkkinenVG,\HenningsonGX,\FukumaBZ}. Note that
\cslikeequation\ is the asymptotic (and therefore local) form of the
CS equation. For finite $r$, \cslikeequation\ becomes non-local.

By varying \cslikeequation\ with respect to $\phi^a({\bf x})$ and
using \realcorrel, we arrive at a local form of the CS equations:
\eqn\csequation{\eqalign{&\[\rho({\bf x}) \frac{\delta}{\delta
\rho({\bf x})} + \sum_b \beta^b({\bf x})
\frac{\delta}{\delta\phi^b({\bf x})}\] \<O_{a_1}({\bf x}_1)\dots
O_{a_n}({\bf
 x}_n)\>_c\cr -&\sum_{k = 1}^n \int d^d x \gamma^{b_k}_{a_k}({\bf x},{\bf x}_k)
    \<O_{a_1}({\bf x}_1)\dots O_{b_k}(\bx)\dots O_{a_n}({\bf
 x}_n)\>_c
    = 0~,}}
where
\eqn\addefinition{
    \gamma^b_a(\bx,\bx_b) = -(4-\Delta_a)\delta^b_a\delta(\bx-\bx_a)
    +\frac{\delta}{\delta \phi_a(\bx_a)} \beta^b(\phi(\bx))\ .
}
This definition of the anomalous dimension gives the deviation of
the operator dimension from that at the UV fixed point, rather than
the deviation from that of a free scalar.\foot{Note that in
\refs{\deBoerXF,\deBoerCZ}, the term $4 - \Delta$ is missing from
$\gamma$. It comes from commuting $2g^{\mu\nu}{\delta\over\delta
g^{\mu\nu}}$ through ${\rho^\Delta\over\sqrt
g}{\delta\over\delta\phi}$.}

As pointed out in \refs{\ErdmengerJA}, \csequation\ can be thought
of as the Callan-Symanzik equation for spacetime-dependent
couplings, first described by Osborn \refs{\OsbornGM}.  The beta
functions $\beta^a$ are the coefficients of the trace of the stress
tensor: $\Theta(\bx) = - \sum_a \beta^a(\bx) \CO_a(\bx)$. We can
transform this equation into one which describes the behavior of
$\Gamma_n$ as one rescales all of its arguments ${\bf x}_i$. The
essential point is that if a function $f$ which itself has mass
dimension $\Delta$ is a function of $n$ variables with definite mass
dimension, an infinitesimal rescaling of any $k$ of these variables
can be traded for an infinitesimal rescaling of the other $n-k$
variables plus an overall rescaling of $f$. This statement holds in
the presence of additional constraints on the variables, so long as
one imposes the constraints at the end. Write the $n$-point function
in \realcorrel\ as:
\eqn\scalingform{ \eqalign{
    \Gamma_n & \equiv \langle \CO_{a_1}({\bf x}_1)\ldots \CO_{a_n}({\bf x}_n)\rangle \cr
        & = \Gamma_n({\bf x}_i, \rho(\bx),\tg_{\mu\nu}(\bx), \phi(\bx)) \cr
}}
where $\Gamma$ is taken to be a nonlocal functional of $\rho$,$\tg$, and $\phi$.  Note that
there are additional dimensionful parameters implied by the spacetime dependence of
$\rho$, $\tg$, and $\phi$.  Therefore we can relate a rescaling of $\rho$ itself
for a rescaling of ${\bf x}_i$ and the variation of the $\rho$,$\tg$, and $\phi$ under a rescaling
of the coordinates they depend on.

In the end, if we integrate \csequation\ over ${\bf x}$ and apply the arguments above, we find
that:
\eqn\csequationtwo{\eqalign{ &\sum_{k=1}^n {\bf x}_k \cdot
\frac{\p}{\p {\bf x}_k}\<O_{a_1}({\bf x}_1)\dots O_{a_n}({\bf
x}_n)\>_c\cr
    &\ \ \ \ \ + \int d^4 y \[ \sum_b \left(-\by\cdot \frac{\p}{\p\by} \phi^b(\by) + \beta^b(\phi)\right)
        \frac{\delta}{\delta\phi^b(\by)}\] \<O_{a_1}({\bf x}_1)\dots O_{a_n}({\bf x}_n)\>_c\cr
    &\ \ \ \ \  - \int d^4 y\ \[ \left({\bf y}\cdot \frac{\p}{\p {\bf y}} \rho({\bf y})\right)
            \frac{\delta}{\delta\rho(\by)} + \delta \tg_{\mu\nu}({\bf y})
            \frac{\delta}{\delta \tg_{\mu\nu}({\bf y})}\]
     \<O_{a_1}({\bf x}_1)\dots O_{a_n}({\bf x}_n)\>_c\cr
     +&\sum_{k = 1}^n \(\delta_{a_k}^{b_k}\Delta_{a_k}-\tilde\gamma^{b_k}_{a_k}(\phi)
    \)\<O_{a_1}({\bf x}_1)\dots O_{b_k}({\bf x}_k)\dots O_{a_n}({\bf x}_n)\>_c
    = 0\ .}}
where
\eqn\constantad{
    \tilde\gamma^b_a = - (4 - \Delta_a)\delta^b_a + \nabla_a \beta^b
}
is the anomalous dimension matrix, and the variation of $\tg$ under
rescaling is:
\eqn\metricvar{
    \delta\tg_{\mu\nu} = \bx\cdot\p_\bx \tg_{\mu\nu} - x^{\lambda}\p_{\mu}\tg_{\lambda\mu}      - x^{\lambda}\p_{\nu}\tg_{\mu\lambda}\ .
}
If $\phi$ and $\rho$ are constant, Eq. \csequationtwo\ is the
standard form of the renormalization group equations, which follows
directly from the Ward identities for broken scale invariance (see
for example \refs{\FreedmanWX,\ZamolodchikovTI}.) The beta functions
are the coefficients of the trace of the stress tensor:
\hbox{$\Theta(\bx) = - \sum_a \rho^{\Delta_{-,a}}\beta^a(\bx)
\CO_a(\bx)$} (where the factor of $\rho$ ensures that $\Theta$ has
dimension $d$.)  In \S5.2, we will show by direct calculation in the
perturbed CFT that the additional terms in \csequationtwo\ for
spacetime dependent $\phi,\rho$ in \csequationtwo\ must also appear.

\subsubsec{Caveats}

We close this section with two comments.  First, in discussions of
the holographic renormalization group
\refs{\BalasubramanianJD\deBoerXF-\deBoerCZ}, it is assumed that
\realcorrel\ at finite $z$ is a good definition of the correlation
functions in a theory cut off at the energy scale $\Lambda = 1/z$.
This statement requires some care. As we will discuss in \S5.5,
there is not a direct relationship between $\phi({\bf x},z_{UV})$
and the coupling $\lambda({\bf x})$ at finite $z_{UV}$: rather,
$\phi$ is determined by both the coupling and the state of the
theory. In the Euclidean calculations,
$\phi$ will be a function of the couplings; we can therefore imagine
a scheme in which $\phi$ {\it  is}\ the coupling, although this may
be related to more standard schemes by a complicated and possibly
nonanalytic redefinition of the couplings.

Secondly, the choice of solution to the Hamilton-Jacobi equations
for the right hand side of \realcorrel\ changes the interpretation. In
general the solution $S$ is a function of constants of motion ${\bf
a}$ (as written in \fivedhjsolution) .  When one takes the
derivatives in \correl, one is varying over classical solutions with
the constants ${\bf a}$ held fixed.  The result depends on the
definition as well as the numerical value of the
constants of motion. On the other hand, the
Euclidean correlators should be uniquely determined by the
couplings. The assumption here, and in other works on holographic
renormalization, is that the right hand side of \realcorrel\ should
be the nonlocal part of $S_{reg}$ in \cutoffact.

\newsec{Holographic RG in Lorentzian signature}

In this section we will generalize the results of
\refs{\deBoerXF,\deBoerCZ}\ and of \S4\ to the Lorentzian version of
the AdS/CFT correspondence. The essential difference is that
normalizable, nonsingular solutions to the equations of motion exist, so that the
coefficient $\lambda^a$ defining the asymptotic behavior $\phi^a
\to_{z\to 0} z^{\Delta_{-,a}} \lambda^a$ no longer unambiguously defines
solutions to the equations of motion. The dual of this statement is
that correlation functions in the perturbed CFT will depend on both
the couplings and the state of the field theory \refs{\BalasubramanianDE,\BalasubramanianSN}.

This leads to our solution to the main question posed in the
introduction. Solving the second-order equations of motion for
supergravity fields via the Hamilton-Jacobi method involves two
steps. The first step is to solve the Hamilton-Jacobi equations. The solutions
are functions of the values of the fields at fixed $z$; the functional form will
depend on the state of the system. The second step
is to solve (the first-order) Hamilton's equations, which requires
that one specifies the values of the fields at fixed $z$. In the
field theory dual, the first step is dual to solving the
Callan-Symanzik equations, and to computing the beta functions. We
will see that the Callan-Symanzik equations and their solutions are
modified by the choice of state in precisely the same way that the
Hamilton-Jacobi equation is.  The second step is roughly dual to
solving the first-order equations $\Lambda\p_{\Lambda} \lambda^a =
\beta^a$ specifying the flow of the couplings with scale. (We say
roughly because the precise duality defines a flow of a combination
of the couplings and the one-point functions of the associated
operators, as we will discuss.)

Thus, despite the apparent first-order nature of the RG equations,
they contain all of the information needed to specify any desired
solution to the dual supergravity equations of motion.  In
\S5.1-\S5.3\ we show this in detail in the limit $z\to 0$. In \S5.4\
we discuss a different class of solutions to the Hamilton-Jacobi
equations and discuss their interpretation in terms of a quantum
field theory with an IR cutoff. In \S5.5\ we discuss the extension
of holographic RG deep into the infrared of the field theory. In
\S5.6\ we discuss the bulk picture of Wilsonian renormalization.

\subsec{The Hamilton-Jacobi functional in Lorentzian signature}

As in \S4.2, the spacetime equations of motion can be solved by
specifying the fields at some constant radial coordinate -- $r$ in
\globalcoords\ and $z$ in \poincoordinates\ -- and solving the radial
Hamilton-Jacobi equations. Any solution to the Hamilton-Jacobi
equation $S$, corresponds to a choice of some constant of motion
$\ba$. As in \S4, one can try to solve the HJ equation with the
on-shell action $S_{SUGRA}$. However, for a general constant of
motion $\ba$, by varying $S_{SUGRA}$ along the set of classical
solutions with fixed $\ba$, the expression $\pi(\phi) = \frac{\delta
S}{\delta \phi}$ for the classical momentum fails due to boundary
terms at $\Sigma_{\pm}$ (see Figs. 1 and 2).\foot{This boundary is a
spacelike slice at any $r$($z$) in global (Poincar\'e) coordinates. In Poincar\'e
coordinates, in the limit $t_{\pm}\to\pm\infty$,
these boundaries include the $z\to\infty$ surface $H_- \cup H_+$ as
well.} $S_{SUGRA}$ will solve the HJ equation if and only if there
is a choice of constant of motion $\ba$ for which the variation of
$S_{SUGRA}$ do not produce boundary terms at $\Sigma_{\pm}$. As
explained in \S3, that is exactly the case when the state is held
fixed. We therefore conclude that the on-shell SUGRA action cut off at
some distance from the boundary
 \eqn\regulatedLorentz{
    S(\phi_{UV},g^{\mu\nu}_{UV}, r_{UV};\ba) = \int^{r_{UV}} dr d^4 x \sqrt{g}\CL(\phi,g^{\mu\nu})~,}
evaluated on a family of classical solutions interpolating between fixed initial and
final states, solves the bulk
radial Hamilton-Jacobi equation.  Furthermore, \regulatedLorentz\ generates the
boundary correlation functions cut off at some scale $l(r_{UV})$:
\eqn\correlators{
    \frac{1}{\sqrt{g({\bf x}_1)}}
    \frac{\delta}{\delta\lambda^{a_1}(\bx_1)} \cdots \frac{1}{\sqrt{g({\bf x}_n)}}
    \frac{\delta}{\delta\lambda^{a_n}(\bx_n)}
        S = \bra{\psi_+} T\[\CO_{a_1}(\bx_1)\ldots \CO_{a_n}(\bx_n)\]
            \ket{\psi_-}_c~.
}
Here $\psi_{\pm}$ denote the eigenvalues of the scalar fields and the metric
at times $t_{\pm}$, and the subscript $c$ denotes the connected correlation functions.
In these equations $\ba$ represents the data which specifies the eigenstates
of the bulk fields at $\Sigma_{\pm}$. $\lambda^a({\bf x})$ are the dimensionful couplings.
In \S5.5 we will argue that the relationship between $\phi$ and $\lambda$ are
nontrivial; near the boundary of AdS, however, they will be related by
a power of the scale factor of the metric, as in \S4.

The constant of motion $\bb$ is given by \bconst. Since the
classical solution is uniquely determined by the initial and final states and the
UV couplings \MarolfFY, the constant of motion $\bb$ is a function
of the states and the UV coupling (and does not depend on the radial
coordinate.)

We are now close to a solution of problem (1) in \S1. Next, in
\S5.2\ we will write the Callan-Symanzik equation for matrix
elements of time-ordered products between arbitrary states. Finally,
in \S5.3\ we will show that the same Callan-Symanzik equations
arises as part of the Hamilton-Jacobi equations for $S$. This will
complete our solution.

For the remainder of this section we will stick to Poincar\'e
coordinates for simplicity's sake.

\subsec{The Callan-Symanzik equation for nontrivial matrix elements}

As we have just intimated, the bulk data which completely specify a solution to the
supergravity equations of motion are dual to the couplings and state of the
field theory.  The non-vacuum states are the new ingredient in the Lorentzian
version of the correspondence.  To understand holographic renormalization,
we must therefore derive the Callan-Symanzik equation for general matrix elements of
time-ordered products of operators:\foot{We would like to thank C. Beasley and H. Schnitzer for
discussions about this derivation.}
\eqn\matrixelements{
    C_{n;\pm} = \bra{\psi_+(t_+)}T\left[\CO_1(\bx_1)\ldots \CO_n(\bx_n)\right] \ket{\psi_-(t_-)}~,
}
for $d$-dimensional conformal field theories perturbed by the interaction Hamiltonian
\eqn\interactlagrang{
    S_{int} = \sum_a \int_{t = t_-}^{t = t_+}
    d^d x \eps^{\Delta_a-d} u^a(\bx) \CO_a(\bx)~,
}
where $\CO_a$ are marginal and relevant operators of dimension
$\Delta_a$, $\eps$ is a cutoff scale with dimensions of length, and
$u^a$ are spacetime-dependent dimensionless couplings (as in
\refs{\OsbornGM}.) We will assume that the operators $\CO_k$ in
\matrixelements\ are scalar operators of definite dimension
$\Delta_k$ at the UV fixed point and that the background is flat
$d$-dimensional Minkowski space. In general, however, the spacetime
dependence of the couplings in \interactlagrang\ means that
couplings to nonscalar operators will be generated along the RG
flow. The sum in \interactlagrang\ should be taken to include these
couplings.  A more elegant treatment would be to consider the
couplings to higher-spin operators as background gauge, metric, and
tensor fields after the fashion of \refs{\OsbornGM}.  We leave this
for future work.

The starting point is the statement that:
\eqn\correlatorderivatives{
\eqalign{
    & \sum_k \left( \bx_k \cdot \p_{\bx_k} + \hat{D}_k\right)
    \bra{\psi_+}T\left(\CO_1(\bx_1)\ldots\CO_k(\bx_k)
        \ldots\CO_n(\bx_n)\right)\ket{\psi_-}\cr
    & \ \ \ \ \ \ \ = i \sum_k \bra{\psi_+}T\left( \CO_1(\bx_1) \ldots [Q(t_k),\CO_k(\bx_k)]\ldots
    \CO_n(\bx_n)
        \right)\ket{\psi_-}
}}
Here $Q(t_k) = \int J^0$ is the charge corresponding to the scale
current $J^{\nu} = x^{\mu}T_\mu^\nu$:
\eqn\scalecurrent{
    Q(t_k) = \int_{t = t_k} d^{d-1} x x^{\mu} T_\mu^0
}
with $T_{\mu\nu}$ the stress tensor. $\hat{D}$ is the dilatation operator defined by
\eqn\dilatationoperator{
    \CO_a(\bx + \lambda\bx) = \CO_a(\bx) + \lambda \bx\cdot \p_\bx \CO_a + \lambda \hat{D} \CO_a(\bx)+\dots~,
}
and the subscript on $\hat{D}$ in \correlatorderivatives\
indicates which of the $\CO_a$ it acts on.

Now
\eqn\gausslaw{
\eqalign{
    & \bra{\psi_+} T(\CO_1\ldots (\CO_k Q(t_k) \CO_{k+1} - \CO_k Q(t_{k+1})\CO_{k+1})
        \ldots\CO_n)\ket{\psi_-} \cr
        &\ \ \ \ \ \ =
        \int_{t_{k+1}}^{t_k} \bra{\psi_+} T( \CO_1\ldots \CO_k \p_0 Q(t) \CO_{k+1}\ldots
            \CO_n ) \ket{\psi_-}\ ,
}}
which we can combine with \correlatorderivatives\ to write:
\eqn\finalwi{
\eqalign{
    & \sum_k \left( \bx_k\cdot\p_{\bx_k} + \hat{D}_k\right)
    \bra{\psi_+}T\left(\CO_1(\bx_1)\ldots\CO_k(\bx_k)
        \ldots\CO_n(\bx_n)\right)\ket{\psi_-}\cr
    & = - i \int_{t_-}^{t_+} dt \bra{\psi_+}
        T\left(\p_t Q(t) \CO_1\ldots \CO_n\right) \ket{\psi_-}\cr
    & \ \ \ \ \ \ \ + i \bra{\psi_+}Q(t_+) T\left(\CO_1\ldots \CO_n\right)\ket{\psi_-}
        - i \bra{\psi_+} T\left(\CO_1\ldots \CO_n\right)
        Q(t_-)\ket{\psi_-}~.
}}
Note the presence of the two extra boundary terms in the last line. These
vanish when  $\ket{\psi_-} = \ket{\psi_+} = \ket{0}$, where $\ket{0}$ is the scale-invariant vacuum state.

We can rewrite $\int dt \p_t Q(t) =  \int d^d x \p_t (x^{\mu}T_{\mu}{}^0)$ via integration by parts:
\eqn\scaletotrace{
\eqalign{
    \int d^d x \p_t (x^{\mu} T_{\mu}{}^0) & = \int d^d x \left(T_{0}{}^0 + x^{\mu}\p_0 T_{\mu}{}^0 \right)
        = \int d^d x \left(T_0^0 - x^{\mu}\p_i T_{\mu}{}^i + x^{\mu}\p_{\nu}T_{\mu}{}^{\nu}\right)\cr
        & = \int d^d x \left( T_0^0 + \p_i (x^{\mu}) T_\mu^i + x^{\mu}\p_{\nu}T_{\mu}{}^{\nu}\right)
            = \int d^d x \left( T^{\mu}_{\mu} + x^{\mu}\p_{\nu}T_{\mu}{}^{\nu} \right) \cr
            & \equiv \int d^d x \left( \Theta +
            x^{\mu}\p_{\nu}T_{\mu}{}^{\nu}\right)~,
}}
where we have assumed that the scale current vanishes at spatial infinity,
or that the spatial directions have no boundary.
The spacetime dependence of the couplings $u$ implies
that the stress tensor is {\it not} conserved; thus, the last equation will not
in general vanish.

Now, we set\foot{The factor of the cutoff is needed for $\Theta$ to
have dimension $d$.} $\Theta(\bx) = - \beta^a(\bx) \CO_a(\bx)
\eps^{\Delta_a-d}$, $\p_a \equiv \frac{\p}{\p u^a(\bx)}$, and
$\p_a(\bx) \CO_k(\by) = B_{ak}^b(\bx,\by) \CO_b(\by)$, to find that:
\eqn\tracetoderivative{
\eqalign{
    & \int d^d x \bra{\psi_+} T\left(\Theta(\bx) \CO_1 \ldots \CO_n\right)\ket{\psi_-} \cr
    =- &i \int d^d x \beta^a \[\p_a \bra{\psi} T(\CO_1 \ldots \CO_n)\ket{\psi_-}
    +\sum_k B^a_{k b}(\bx,\bx_k) \bra{\psi_+} T(\CO_1\ldots
    \CO_k\ldots \CO_n) \ket{\psi_-}\]\cr
    -&\int_{t_+} d^{d-1} x \beta^b \bra{\psi_+} \hat{B}_b T(\CO_1\ldots \CO_n) \ket{\psi_-}
    + \int_{t_-} d^{d-1} x \bra{\psi_+} T(\CO_1 \ldots \CO_n) \hat{B}_b \ket{\psi_-}
        \beta^b~,
}}
where $i \p_b \ket{\psi} = \hat{B}_b \ket{\psi}$, $-i\p_b \bra{\psi} = \bra{\psi} \hat{B}$.
With the factor of $i$, $\hat{B}$ is Hermitian.

Finally, if we define:
\eqn\scalingmatrices{
\eqalign{
    \hat{D}_a \CO_a & = \Gamma_a^k \CO_k\cr
    \gamma^a_k(\bx,\by) & = \Gamma^a_k\delta(\bx-\by) - \beta^b B^a_{kb}(\bx,\by)\cr
    \hat{K}(t)  & = \left[Q(t) + \int_t d^{d-1} x\beta^b \hat{B}_b\right]
}}
then the Callan-Symanzik equation is:
\eqn\modifiedCS{
\eqalign{
    & \sum_k  \bx_k\cdot\d_{\bx_k} \bra{\psi_+}T(\CO_1\ldots\CO_n)\ket{\psi_-}
    + \sum_k \int d^dx\ \gamma^a_k(\bx,\bx_k) \bra{\psi_+} T(\CO_1\ldots\CO_a(\bx_k)\ldots\CO_n)\ket{\psi_-}\cr
    &\ \ \ \ \ + \int d^dx\[\beta^a \p_a \bra{\psi_+}T(\CO_1\ldots\CO_n)\ket{\psi_-}
    +  \bra{\psi_+} T\left( x^{\mu}\p_{\nu}T_{\mu}{}^{\nu}(\bx) \CO_1
        \ldots \CO_n\right)\ket{\psi_-}\] \cr
    &\ \ \ \ \ = i  \bra{\psi_+}\hat{K}(t_+) T(\CO_1\ldots\CO_n)\ket{\psi_-}
    - i \bra{\psi_+}T(\CO_1\ldots\CO_n)\hat{K}(t_-)\ket{\psi_-}~.
}}
The last term on the second line is the local modification of the
Callan-Symanzik equation due to the spacetime-dependent couplings.
The two boundary terms on the final line are the modification of the
Callan-Symanzik equation for general matrix elements of time-ordered
products of operators.\foot{If we define the states as
integrals of local operators acting on the vacuum, then these last
two terms can be derived from the Callan-Symanzik equation for
vacuum correlators.}

To match \csequationtwo\ more precisely, let us consider the contribution of the scalar operators
in \interactlagrang\ to $x^{\mu}\p_{\nu}T_{\mu}{}^{\nu}$.
The arguments of Noether's theorem applied to the perturbed CFT leads to the equation
\eqn\newdivergence{
    x^{\mu}\p_{\nu} T_{\mu}{}^{\nu}(\bx) = - \sum_a \eps^{\Delta - d} (\bx\cdot \p_\bx u^a(\bx)) \CO_a(\bx)\ .
}

Now, if we define
\eqn\newbeta{
    \tilde{\beta}^a(\bx) = \beta^a(\bx) - \bx\cdot \p_\bx u^a(\bx)~,
}
for the scalar operators, and replace $\beta$ with $\tilde{\beta}$ in \scalingmatrices, then we can write
the Callan-Symanzik equation in the form:
\eqn\anothercs{
\eqalign{
    & \sum_k  \bx_k\cdot\p_{\bx_k} \bra{\psi_+}T(\CO_1\ldots\CO_n)\ket{\psi_-}
    + \sum_k \int d^d x\ \gamma^a_k(\bx,\bx_k) \bra{\psi_+} T(\CO_1\ldots\CO_a(\bx_k)\ldots\CO_n)\ket{\psi_-}\cr
    &\ \ \ \ \ + \int d^d x \tilde{\beta}^a \p_a \bra{\psi_+}T(\CO_1\ldots\CO_n)\ket{\psi_-}
        + \int d^d x \bra{\psi_+} T\left( x^{\mu}\p_{\nu}\delta
        T_{\mu}{}^{\nu}(\bx)
            \CO_1\ldots \CO_n\right)\ket{\psi_-}\cr
    &\ \ \ \ \ = i  \bra{\psi_+}\hat{K}(t_+) T(\CO_1\ldots\CO_n)\ket{\psi_-}
    - i \bra{\psi_+}T(\CO_1\ldots\CO_n)\hat{K}(t_-)\ket{\psi_-}\ ,
}}
where $\delta T$ is the contribution of the non-scalar operators.
Note that the shift from $\beta$ to $\tilde{\beta}$ is precisely what we find in the second line
of \csequationtwo.

\subsec{Lorentzian HJ/CS correspondence}

Next, we must understand how the Hamilton-Jacobi equations
are related to the Callan-Symanzik equations. To do so,
we will carry over the strategy of \S4.3\ for solving the
Hamilton-Jacobi equations. The major difference is that we must
include the dependence on the states.

As in \S4, the SUGRA action can be written in a derivative
expansion. The regularized generating function $\Gamma$ is then
obtained by the subtraction of some local counterterms $S^{(0)}$ and $S^{(2)}$:
$\Gamma=S-(S^{(0)}+S^{(2)})$. Apart from the signature difference,
the functional form of the bulk Lagrangian is the same as the
Euclidean one.  For general states there may be additional
boundary terms $\psi_\pm$ at $\Sigma_\pm$ in the supergravity action
\MarolfFY. However, if we work with eigenstates of the supergravity field operators at $\Sigma_{\pm}$,
these boundary terms will not be present.
Lorentzian signature adds no additional ambiguities to the
solutions of the first two equations in \gradedsolutions.
Therefore, the functional form of the local counterterms $S^{(0)},S^{(2)}$
\actionbreakup\ are the same as in the Euclidean case.
Furthermore, the functional differential equation in the
last line of \actionbreakup\ is the same.
All of the dependence on the constant of motion
$\ba$ (and therefore on the state) is contained in the choice of solutions
$\Gamma$ to \actionbreakup. The field theory dual of this statement is that near the UV fixed point,
the local beta functions are independent of the state, while the correlation functions
depend strongly on the states, especially when the operators are widely separated.

To relate the modified Callan-Symanzik equation \anothercs\ to the
HJ equation, we should adapt the derivation of \csequationtwo\
to the case of Lorentzian signature. Let us write the
boundary coordinates as ${\bf x} = (t,\vec{x})$, where $t$ is the time direction.
The correlation function written as a function of all of the dimensionful parameters of
the problem is:
\eqn\Lorentzianscalingform{
\eqalign{
    \Gamma_n & \equiv\bra{\psi_+(t_+)}T(\CO_1({\bf x}_1)
    \ldots\CO_n({\bf x}_n))\ket{\psi_-(t_-)}_c\cr
    & = \Gamma_n\left({\bf x}_i,\rho(\bx),{\tilde g}_{\mu\nu}(\bx), \phi(\bx),\psi_+, t_+,\psi_-,t_-\right)
 }}
Using the same strategy used in \S4.4, we find that:
\eqn\lorentzcsequation{\eqalign{
&\sum_{k=1}^n {\bf x}_k \cdot \frac{\p}{\p {\bf x}_k}\bra{\psi_+}
    O_{a_1}({\bf x}_1)\dots O_{a_n}({\bf x}_n)\ket{\psi_-}_{c}\cr
    &\ \ \ \ \ + \int d^d y \[ \sum_b \left(-\by\cdot \frac{\p}{\p\by} \phi^b(\by) + \beta^b(\phi)\right)
        \frac{\delta}{\delta\phi^b(\by)}\] \bra{\psi_+}O_{a_1}({\bf x}_1)\dots O_{a_n}({\bf x}_n)
            \ket{\psi_-}_{c}\cr
    &\ \ \ \ \  - \int d^d y\ \[ \left({\bf y}\cdot \frac{\p}{\p {\bf y}} \rho({\bf y})\right)
            \frac{\delta}{\delta\rho(\by)} + \delta\tg_{\mu\nu}
            \frac{\delta}{\delta \tg_{\mu\nu}({\bf y})}\]
    \bra{\psi_+}O_{a_1}({\bf x}_1)\dots O_{a_n}(\bx_n)\ket{\psi_-}_{c}\cr
     &\ \ \ \ \ +\sum_{n = 1}^k \(\delta_{a_k}^{b_k}\Delta_{a_k}-\tilde\gamma^{b_k}_{a_k}(\phi)
    \)\bra{\psi_+}O_{a_1}({\bf x}_1)\dots O_{b_k}({\bf x}_k)\dots O_{a_n}({\bf x}_n)\ket{\psi_-}_{c}\cr\cr
    & =  \left[ \int d^{d-1} \vec{y} dz \left\{\left[ \left(\vec{y}\cdot\frac{\p}{\p \vec{y}}
    + z\p_z \right)\psi_- \right]\frac{\delta}{\delta\psi_-}
    + \left[\left(\vec{y}\cdot\frac{\p}{\p \vec{y}}
    + z\p_z \right)\psi_+\right] \frac{\delta}{\delta\psi_+}\right\}\right.
    \cr
    &\ \ \ \ \ \ \ \ \ \left. - t_-\frac{\p}{\p t_-} - t_+ \frac{\p}{\p t_+} \right]
    \bra{\psi_+}O_{a_1}({\bf x}_1)\dots O_{a_n}({\bf
    x}_n)\ket{\psi_-}_{c}~.
    }}

The first four lines of this equation are as before. We claim that
the final two lines in \lorentzcsequation\ should be precisely dual
to the large-N limit of the final line in \modifiedCS,\anothercs. To
see this, let us consider the supergravity dual of $\hat
K(t_-)\ket{\psi_-}$.  In the bulk, the state is specified by a set
of functions $\phi^a_-(\vec{x},z)$ where $z$ is the radial
coordinate along $\Sigma_-$.  These functions are the eigenstates of
the field operators. One may decompose these field operators locally
into modes labelled by the ${\vec{x}}$-momentum ${\bf k}$, and the
{\it frequency} $\omega$.  In the bulk, the equations of motion
relate the $\vec{k}$, $\omega$-dependence to the $\vec{x},z$
dependence.  This map has been made explicit in the large-N limit of
unperturbed AdS spacetimes in \refs{\BalasubramanianDE,\BanksDD}. In
general all we need is that such a map exists, and that the creation
operators of the bulk fields are a function of the Fourier modes of
the boundary operators.

Now on the boundary, $\CP_x \equiv \int d^{d-1}x\ x^i T_i^0$
generates rescalings of the spatial coordinates $\vec{x}$. It also
rescales the Hamiltonian: the Hamiltonian is an operator with mass
dimension $1$, implying that $\[Q(t_-) +\CP_\lambda(t_-),H(t_-)\] =
i H(t_-)$, where $\CP_\lambda\equiv i\int d^{d-1}x\ \beta^a\d_a$
rescales the dimensionful couplings in $H$ \TreimanEP.\foot{We would like to thank J. McGreevy for pointing out a mistake at this point in the previous version of this paper.}
Since $H$ commutes with itself, this implies that
$\[\CP_x+\CP_\lambda,H\]= iH$. Now, $\ket{\psi_-}$ is an eigenstate
of some function of the Fourier modes of the local operators of the
boundary theory. The frequency of these operators is defined by the
equation
\eqn\commutator{
    \[H,\CO_{\omega}\] = i \omega\CO_{\omega}\ ,
}
and the momenta by their $x$-dependence. The result is that
$(\CP_x+\CP_\lambda)$ rescales all of the four-momenta of boundary
operators. The map to the bulk implies that
$(\CP_x+\CP_\lambda)\ket{\psi_-}$ acts via the penultimate line in
\lorentzcsequation.

Finally, we can use \transamp\ to note that
\eqn\Htotime{
    \bra{\psi_+(t_+)} T \exp\left(\frac{i}{\hbar} \int_{t = t_-}^{t = t_+}
    d^d x \sum_a \lambda^a(\bx) \CO_a(\bx)\right)
        t_-H(t_-)\ket{\psi_-(t_-)} = t_-d_{t_-} Z = t_- \p_{t_-} Z~.
 }
Therefore, if the duality holds, the explicit time derivatives in
the last line of \lorentzcsequation\ should map to the contributions
from $t_{\pm} H(t_{\pm})$ in the final line of \modifiedCS.

To see how the bulk and boundary rescalings of the states map to each other,
consider the example of the
large-N limit of the unperturbed CFT, in the approximation that we can ignore
bulk interactions, and as the cutoff is taken to zero.
Let the state be a coherent classical state in which a single
local, single-trace operator $\CO$ has a macroscopic expectation value:
\eqn\singleev{
    \bra{\phi_+}\CO({\bf x})\ket{\phi_-} = \widetilde\phi({\bf x})= \Gamma_1({\bf x})\ ,
}
In the unperturbed theory, the spacetime equations of motion
imply that $\widetilde\phi(x)$ is a linear functional of $\phi_+$, $\phi_-$, and
has dimension $\Delta_+$. This means that we can write:
\eqn\btobfreetransform{
    \widetilde{\phi}({\bf x}) = \sum_{\eps = \pm} \int _{t_{\eps}}d^{d-1}\vec{y} dz
        \ F_{\eps} ({\bf x}; \vec{y}, z, t_+,t_-) \phi_{\eps}(\vec{y}, z)
}
where $G$ is a function of dimension $\Delta + d$.
Therefore, dimensional analysis implies that the final line in \lorentzcsequation\
acting on $\Gamma_1$ is:
\eqn\ictoev{
\eqalign{
    & \left[ \int d^{d-1} \vec{y} dz \left\{\left[ \left(\vec{y}\cdot\frac{\p}{\p \vec{y}}
    + z\p_z \right)\phi_- \right]\frac{\delta}{\delta\phi_-}
    + \left[\left(\vec{y}\cdot\frac{\p}{\p \vec{y}}
    + z\p_z \right)\phi_+\right] \frac{\delta}{\delta\phi_+}\right\} \right.\cr
    &\ \ \ \ \ \ \ \ \ \ \ \ \left. - t_- \frac{\p}{\p t_-} - t_+ \frac{\p}{\p t_+} \right]
    \widetilde\phi({\bf x}; \phi_+,t_+,\phi_-,t_-)
    = \left({\bf x}\cdot\frac{\vec\p}{\p{\bf x}} + \Delta\right)\widetilde\phi\ .
}}
The required that we rescaled both the arguments of $\phi_{\pm}$ as well as $t_{\pm}$.

Now, let us compare this to the right hand side of \modifiedCS.
Conformal invariance implies that $Q(t)$ is constant in time, so
that we can write:
\eqn\onepointstatevar{
\eqalign{
    & i \bra{\phi_+}Q(t_+)\CO({\bf x})\ket{\phi_-} - i
    \bra{\phi_+}\CO(t,x)Q(t_-)\ket{\phi_-}\cr
    &\ \ \ \ \ \ = i \bra{\phi_+}[Q,\CO({\bf x})]\ket{\phi_-}
    = \bra{\phi_+}\left({\bf x}\cdot \frac{\p}{\p {\bf x}} + \Delta\right)\CO({\bf x})
        \ket{\phi_-}
    = \left({\bf x}\cdot\frac{\p}{\p{\bf x}} +
    \Delta\right)\widetilde\phi~.
}}
This is precisely the variation of $\widetilde\phi$ that we find in \lorentzcsequation,\ictoev.
Note further that \lorentzcsequation\ has only been derived for classical states
in the large-N limit.  For such states in the field theory,
the action of $Q$ on the states will be specified by the scale transformation of the
one-point functions.

We have solved the main problem raised in the introduction.  Let us summarize the basic point.
The Hamilton-Jacobi formalism involves two sets of equations.  One set
of equations is Hamilton's equations for the field variables
$\phi(x,z)$, (roughly) $z\p_z \phi = \frac{\delta S}{\delta \phi}$.
Studies of holographic RG in Euclidean signature indicate
that these equations are mapped into the standard
first-order RG flow equations $\Lambda\p_{\Lambda}\lambda^a =
\beta^a$, at least near the timelike boundary of the 5d spacetime.
Given the HJ functional $S$, the solutions are completely
specified by the values of $\phi$ at some UV scale $z$. For $z\to
0$, this data is dual to the UV couplings $\lambda^a$.

However, the spacetime equations of motion are second order (at
large $N$ and low energies.) In Lorentzian signature their solutions
are only uniquely specified after one specifies additional data,
dual to the presence of normalizable modes. The point is that to
solve the spacetime equations of motion one solves the
Hamilton-Jacobi equation for find $S$, and {\it then} solves
Hamilton's equations for $\phi$.   $S$ depends on additional
constants of motion ${\bf a}$. In the discussion above we have found that
${\bf a}$ labels the classical states of the theory.

We have shown that the Hamilton-Jacobi and Callan-Symanzik equations for correlation functions
are identical (in the large-N limit) even in the presence
of classical states.  Therefore the constants of motion ${\bf a}$
in the Hamilton-Jacobi formulation of supergravity are precisely dual
to data required in the field theory, in order to
specify the scaling behavior of correlation functions.
In other words, the field theory contains all the structure of a theory governed by
second order equations of motion in the bulk.  The
scaling behavior of the theory is determined by both the RG flow equations and
the Callan-Symanzik equations.

One caveat is that while the natural RG flow equations in field theory
specify the variation of the couplings with scale, the associated
Hamilton's equations of the supergravity dual specify the flow of $\phi$ with
$z$.  These two statements are only precisely dual as $z\to 0$.  We will discuss
the case of finite $z$ in \S5.5.

\subsec{Perturbed CFTs with an IR cutoff}

We have studied solutions to the Hamilton-Jacobi
equations with constants of motion that specify the (classical) state of the system.
There are other solutions.  We will discuss one class for which the constants of motion are
dual to one-point functions of operators specified at some infrared scale.
These are not the
same: as discussed in \refs{\BalasubramanianDE}, these one-point functions
depend on both the couplings, the state, and the scale.

We consider single free scalar field excitations at energies much smaller then the Planck
scale. For such low energy excitations backreaction is neglectable
and one can treat the scalar field as it was propagating in a fixed
AdS background. In this limit Hamilton's principal function $S[(g_{\mu\nu}({\bf
x}),\phi({\bf x})),{\bf a}]$ reduces to a function of $\(z, \phi({\bf
x}),{\bf a}\)$, where the metric dependence is replaced by
dependence on $z$.  The corresponding HJ equation reduces to a
form similar to \hprinceq, where $t$ is replaced by $z$.

One solution to the HJ equation, analogous to \hjaction, is
 \eqn\cutoffaction{S[\phi({\bf x}),{\bf a},z]
 = \int^{z_0}_z d\tilde z\int d^d x \CL\(\phi({\bf x},\tilde z)\)~,}
where $z_0 > z$, evaluated on a solution to the equation of motion
with boundary conditions $\phi({\bf x},z)=\phi({\bf x})$, ${\bf a} \equiv \phi({\bf
x},z_0)=\phi_0({\bf x})$.  For $z_0$ deep enough in the
bulk, $\phi$ is dominated by the term scaling as $z^{\Delta_+}$.
Therefore, $z_0^{-\Delta_+}\phi_0$ can be interpreted as the
one-point function of the operator at scale $z_0$.  Note, however,
that with this choice for the constants of motion ${\bf a}$ held
fixed, the nonlocal part of $S$ in \cutoffaction\ is no longer the
generating function of correlation functions. In order to keep the
expectation values $\phi_0$ fixed as one varies $\phi(\bx)$, one
must vary the couplings {\it and} the state.

Our simplified Lagrangian is
 \eqn\LL{\CL\(\phi({\bf x},\tilde z)\)=-\half\tilde{z}^{1-d}
 \[(\p_{\tilde{z}}\phi)^2+\eta^{\mu\nu}\d_\mu\phi\d_\nu\phi+{m^2\over z^2}\phi^2\]~.}
Integrating the kinetic term by parts and using the equation of
motion we find that \cutoffaction\ can be written as two boundary
terms:
 \eqn\boundarytermaction{S[\phi({\bf x}),\phi_0({\bf x})]=
 -\half\int d^dx\tilde{z}^{1-d}\phi \p_{\tilde{z}}\phi |_z^{z_0}=
 \half\int d^dx\[ z^{1-d} \phi \p_z \phi- z_0^{1-d} \phi_0\p_{z_0}\phi_0\]~.}

For simplicity, let us consider a scalar field that almost saturates the
Breitenlohner-Freedman bound $R_{AdS}^2m^2\ge -4$
\BreitenlohnerBM, or equivalently $0<\nu\ll 1$.  The solutions to the
equations of motion will have the following leading behavior at small $z\le z_0\ll R_{AdS}$
\eqn\asymptscalar{
    \phi({\bf x},z)=\alpha({\bf x})z^{\Delta_-}(1+O(z^2)+\ldots)
    +\beta({\bf x})z^{\Delta_+}(1+O(z^2)+\ldots)~.}
In this case both independent solutions are normalizable.
One may choose either of $\alpha$,$\beta$ to be dual to the coupling,
with $\Delta_{\mp}$ the
corresponding operator dimension \refs{\BalasubramanianRI}.  These
are related by a Legendre transformation \refs{\KlebanovTB,\HartmanDY}.
We will consider the case that the operator dimension is
$\Delta_+$, and $\alpha$ is dual to the coupling in the field theory.
The discussion should then connect smoothly to one for operators
of higher dimension, for which the term proportional to $\alpha$ is non-normalizable.

To write \boundarytermaction\ as a solution to the Hamilton-Jacobi equations,
we must find $\p_z\phi, \p_{z_0}\phi_0$ as a function of of $\phi,\phi_0$:
\eqn\radialderivativesfinal{
\eqalign{
    z \p_z \phi &\simeq\left[ \frac{\Delta_-z^{-2\nu}}{z^{-2\nu} - z_0^{-2\nu}} +
            \frac{\Delta_+ z^{2\nu}}{z^{2\nu} - z_0^{2\nu}}\right]\phi
            -\left[\frac{\Delta_-z^{\Delta_-}z_0^{-\Delta_+}}{z^{-2\nu}-z_0^{-2\nu}}
            + \frac{\Delta_+ z^{\Delta_+}z_0^{-\Delta_-}}{z^{2\nu}-z_0^{2\nu}}\right]\phi_0\cr
    z_0 \p_{z_0}\phi_0
    &\simeq-\left[\frac{\Delta_- z_0^{-2\nu}}{z^{-2\nu} - z_0^{-2\nu}} +
        \frac{\Delta_+ z_0^{2\nu}}{z^{2\nu}-z_0^{2\nu}}\right] \phi_0
        + \left[\frac{\Delta_- z^{-\Delta_+}z_0^{\Delta_-}}{z^{-2\nu} - z_0^{-2\nu}}
        + \frac{\Delta_+ z^{-\Delta_-}z_0^{\Delta_+}}{z^{2\nu} - z_0^{2\nu}}\right] \phi
}~.}
The classical action \boundarytermaction\ can be written as:
\eqn\adsHJaction{
\eqalign{
    S \simeq\int d^dx\{&\half \left[ \frac{\Delta_-z^{-d-2\nu}}{z^{-2\nu} - z_0^{-2\nu}} +
    \frac{\Delta_+ z^{-d+2\nu}}{z^{2\nu} - z_0^{2\nu}}\right]\phi^2
    -\left[ \frac{\Delta_- (z z_0)^{-\Delta_+}}{z^{-2\nu} - z_0^{-2\nu}}
    + \frac{\Delta_+ (z z_0)^{-\Delta_-}}{z^{2\nu} - z_0^{2\nu}}\right]\phi\phi_0\cr
    +&\half \left[ \frac{\Delta_-z_0^{-d-2\nu}}{z^{-2\nu} - z_0^{-2\nu}} +
    \frac{\Delta_+ z_0^{-d+2\nu}}{z^{2\nu} -
    z_0^{2\nu}}\right]\phi^2_0\}
}~.}
As $z\to 0$ the dimensionless UV coupling becomes
$u=\alpha z^{\Delta_-}$. The dimensionless one-point function of the
dual operator at scale $z_0$ is $\widetilde{u}=\beta z_0^{\Delta_+}$.We take
$u,\widetilde u\sim 1$: the coupling is specified at a UV scale, and the
one-point function at some IR scale. With $u,\widetilde u$ so specified, $\beta $
will dominate over $\alpha$ at $z_0 \gg z$, as
$\beta z_0^{\Delta_+}\gg\alpha z_0^{\Delta_-}$ when
$\widetilde{u}\gg u\({z\over z_0}\)^{\Delta_-}$. If we take the limits
$z_0\ll R_{AdS}$ and $z/z_0\ll 1$, we will find that the
interpretation of flow in $z$ as RG flow is particularly clean. In particular,
the action simplifies:
\eqn\newoolaction{
    S \sim \int d^dx\[\half \Delta_- z^{-d}\phi^2 +2\nu z^{-\Delta_-}
        z_0^{-\Delta_+}\phi\phi_0-\half\Delta_+ z_0^{-d}\phi_0^2\]~.}
This satisfies the Hamilton-Jacobi equation in the limit specified above, by construction.
Next, Hamilton's equations for $\phi$ are:\foot{Note that since $z$ is the lower bound
of the integral \cutoffaction, $\pi_\phi=-{\delta
S\over\delta\phi}$.}
 \eqn\hjnewooleomone{-z^d \pi_\phi=z\p_z \phi =z^d \frac{\delta
S}{\delta\phi}\sim\Delta_- \phi + 2\nu z^{\Delta_+} z_0^{-\Delta_+}
\phi_0~.}
The term $\Delta_- \phi$ is just the beta function to linear order.
The second, subleading term, controls the one-point function fo the dual operator.
The general solution to \hjnewooleomone\ is:
\eqn\hjnewoolsolutionfour{
    \phi({\bf x},z) \sim z^{\Delta_-}\lambda({\bf x}) +
    z^{\Delta_+}z_0^{-\Delta_+}\phi_0({\bf x})~,
}
where $\lambda({\bf x})$ corresponds to the coupling.

In this limit, \cutoffaction\ has the form of the solutions
discussed in \S3.  Here $\half \Delta_-\phi^2$ matches the "local" contribution $S^{(0)}$. The
other two terms in \newoolaction\ belong to $\Gamma$.
Note that
\eqn\firstderiv{
    \p_{\phi}\Gamma|_{\phi_0\  \rm{fixed}} = 2\nu z^{-\Delta_-}
        z_0^{-\Delta_+}\phi_0 = \<\CO\>\ .
}
However, $\p^2_{\phi}\Gamma|_{\phi_0\ {\rm fixed}} = 0$. Because we have kept $\phi_0$
rather than $\widetilde\phi$ in our variations, variations of $\Gamma$ with respect to $\phi$
are not the correlation functions of the theory.

The action \adsHJaction\ has a symmetry under exchanging
$(z_0,\phi_0,\Delta_+)$ with $(z,\phi,\Delta_-)$ and flipping the
overall sign. Thus, taking the opposite limit $z_0/z\to
0$ with $z\ll R_{AdS}$, will result in exchanging the roles of
$\phi_0$ and $\lambda$. This limit matches the discussion of
\S3\ if we choose the second root $\vartheta_a=\Delta_{a,+}$ in \num.

We should note that this solution to the Hamilton-Jacobi equation could
just as easily have been found in Euclidean space.  The fields will be
nonsingular in the region between $z,z_0$.  Only when the IR cutoff is
removed will we be forced to choose a particular value for the normalizable mode.
This statement has an candidate analog in the dual field theory.  Conformal
perturbation theory for relevant perturbations is plagued by infrared
divergences: the proper treatment of these divergences requires
adjusting the one-point functions of the theory
(\cf\ \refs{\WilsonZS\JackiwKV-\ZamolodchikovBK}.)

\subsec{Extension of the holographic RG equations into the IR}

As with other discussions of the holographic RG formalism, ours has taken place deep in the
UV region ($z\to 0$) of the theory.  There are a number of issues with extending
the equations into the IR, some of which become even more difficult in the
Lorentzian description.  Many of these have been mentioned elsewhere, but we
wish to collect the problems here and expand on them.

\vskip .2cm

\noindent \item{{\bf 1.}} The identification of $\phi({\bf x},z)$ as
the dimensionless coupling at some scale $l(r)$ was based on the
asymptotic behavior of $\phi$ as $z\to 0$ in \boundarybeh. For
finite $z$, the relation between $\phi$ and the coupling is more
complicated.  (See also \refs{\KalkkinenVG,\BergHY} for a discussion
of this issue.) First, eq. \classicalfields\ shows that $\phi$ is in
general a sum of contributions from the couplings and contributions
from the state. Since the procedure of integrating out modes will
depend on the properties of the state (near the cutoff), one might
imagine that as one lowers the cutoff (identified with $z$), $\phi$
can be identified with the renormalized coupling in some scheme.
However, it is unclear to us that such a scheme exists and is useful
in the dual field theory.

\item{} \ \ \ \ \ Even without such a scheme, some sort of relationship between the
Hamilton-Jacobi and Callan-Symanzik equations should hold. However,
the simple local, linear relation used in \realcorrel\ will no longer hold.
More generally, $\rho^{-\Delta_-}{\delta\over\delta\phi(\bx,z)}$ in \realcorrel\
should be replaced with
 \eqn\exactvar{\int
 d^dy{\delta\phi(\by,z)\over\delta\widetilde\lambda_z(\bx)}{\delta\over\delta\phi(\by,z)}~,}
where $\tilde\lambda_z$ is the dimensionless coupling at scale $z$, and
the derivative $\delta_{\tilde\lambda_z}\phi$ is taken with the state fixed.

\vskip .2cm

Nonetheless let us continue to discuss the Hamilton-Jacobi equations.
The scaling of the fields identified by \refs{\deBoerXF,\deBoerCZ}\ becomes
complicated at finite $z$, as the authors of those references indeed point out.
This leaves less of a reason to solve the Hamilton-Jacobi
equations by breaking them up as in \gradedsolutions.
Nonetheless, the derivative expansion in the bulk remains valid, and
the first two lines of \gradedsolutions\ still have the same solution as before.
However, there were additional terms in the Hamilton-Jacobi equation
that were dropped in the small-$z$ approximation, that we can no longer drop.
These modify the third equation in \gradedsolutions.
The Hamilton-Jacobi equations become:
\eqn\threepiec{\eqalign{
    & \left\{S^{(0)},S^{(0)}\right\}  = \CL^{(0)} \cr
    & 2 \left\{S^{(0)}, S^{(2)}\right\}  = \CL^{(2)} \cr
    & 2 \left\{S^{(0)}+S^{(2)}+\Gamma,\Gamma\right\}  -  \left\{\Gamma,\Gamma\right\}=
   - \left\{S^{(2)},S^{(2)}\right\} }~.}
where $\Gamma \equiv S - S^{(0)} - S^{(2)}$.  We will assume that $\Gamma$ is the generating
function of correlation functions in the cutoff theory.
The first term in the third equation can be rewritten via the full set of Hamilton's equations
\eqn\phidot{\eqalign{\frac{\p\phi^a({\bf x},r)}{\p r} &
=\frac{G^{ab}(\phi)}{\sqrt{g}}{\delta \over \delta\phi^a({\bf
x},r)}\[S^{(0)}+S^{(2)}+\Gamma\]\cr \frac{\d{g}_{\mu\nu}({\bf
x},r)}{\p r} & = \frac{1}{\sqrt{g}}\(- 2
\frac{\delta}{\delta{g}^{\mu\nu}({\bf x},r)} + \frac{2}{3}
g_{\mu\nu}g^{\lambda\rho}\frac{\delta}
 {\delta{g}^{\lambda\rho}({\bf x},r)}\)\[S^{(0)}+S^{(2)}+\Gamma\]}~,}
such that
 \eqn\genCSprelim{\eqalign{
 \left( - \frac{d}{d r} + \frac{\p}{\p r}\right)
 \Gamma & \equiv \int d^d x \[{\d g_{\mu\nu}\over\d r}g^{\mu\rho}g^{\nu\sigma}
 {\delta\over\delta g^{\rho\sigma}({\bf x},r)}
 -{\d\phi^a\over\d r}{\delta\over\delta\phi^a({\bf x},r)}\]\Gamma\cr
 &\ \ \ \ \ = -\int d^dx \left\{S^{(2)},S^{(2)}\right\} +
 \int d^d \bx\left\{\Gamma,\Gamma\right\}~.}}
Eq. \fivedhjsolution\ states that $\p_r S = 0$.  Since we can see explicitly that
$\p_r (S^{(0)} + S^{(2)} ) = 0$, this implies that $\p_r \Gamma = 0$ as well, leaving us with the
tantalizing equation:
\eqn\genCS{
    d_r \Gamma + \int d^d x  \{\Gamma,\Gamma\} = \int d^d x \{S^{(2)},S^{(2)}\}~.
}
We leave the field-theoretic interpretation of this equation for future work.
Note that without the $\{\Gamma,\Gamma\}$ term, this looks like an integrated form
of Osborn's version of the Callan-Symanzik equations \refs{\OsbornGM}.

There are two further problems with relating \genCS\
to the field theoretic Callan-Symanzik equations.

\vskip .2cm

\item{{\bf 2.}} As discussed in point (1) above, the relation between $\phi$
and the couplings as typically defined may be complicated, and
requires information about the quantum state. Therefore, there is a lot
of work to relate $\p_r\phi \delta_{\phi}\Gamma$ to $\beta\p_{\lambda}\Gamma$.
Note that for spacetime dependent couplings,
one {\it does} expect nonlocal contributions to the beta function
(as mentioned in \refs{\OsbornGM}), so some piece of the
contribution to $\p_r\phi$, $\p_r g$ from $\p_{\phi}\Gamma$ and
$\p_g \Gamma$  may appear in the field-theoretic beta
functions.

\item{{\bf 3.}} We have considered deformations in the UV by single-trace operators only.
However,  multiple-trace operators will generically be induced
under the RG flow \AharonyPA. Nonetheless, consider the (infinite-dimensional)
surface in the space of couplings which is swept out by RG trajectories  which
are purely single-trace in the ultraviolet. For our Hamilton-Jacobi equations
to successfully capture the large-N RG equations, we are assuming that in our scheme,
$\phi^a({\bf x},z)$ are good coordinates on this surface.

\item{\bf 4.} The role of the term $\{\Gamma,\Gamma\}$ on the left
hand side of \genCS\ is not understood.

\vskip .2cm

We leave these issues for future work.

\subsec{Reversibility of holographic RG}

The work of Susskind and Witten \refs{\SusskindDQ}\ suggests that
"cutting off" the asymptotic region $z < \epsilon$ of AdS space is
dual to a spatial cutoff in the dual field theory. However,
Wilsonian renormalization, achieved by tracing out degrees of
freedom at scales larger than the cutoff, is not a reversible
process, while the second order supergravity equations can be
integrated either out towards the boundary or in towards the
interior. If $S_{reg}$ in \cutoffact\ or \regulatedLorentz\ is meant
to describe the quantum field theory cut off at distance scale
$z_{UV}$, why then can we integrate the equations of motion out to
the boundary?

In the classical limit of the spacetime theory, the answer is that
the cutoff in $S_{reg}$ merely smooths out the short-distance
singularities of the theory, and does not set them to zero. For
example, we can see that the two-point function is smooth and
generally nonvanishing as the separation vanishes.\foot{The point
that the finite-$z$ cutoff is a complicated "smearing" function has
been made, for example, in \refs{\BalasubramanianJD}.}  In our
discussion until this point, no limitation has been placed on the
sensitivity of our measurements, so that we can specify information
about the theory at all scales, even in the presence of a cutoff. If
this information includes all possible irrelevant operators (dual to
massive fields in the bulk spacetime), then we can follow the theory
into the UV without any obstruction.

In practice, detectors sensitive to gauge theory observables
will have limited accuracy. The detectors could have finite spatial resolution
$\ell$, or they could have finite sensitivity to the amplitude of
the fluctations.  More generally both limitations will be in effect.
In the former case, one would naturally perform experiments with
$z_{UV}$ set equal to $\ell$; the finite resolution makes it
impossible to follow the theory into the UV.  In the second case, at
any given cutoff $z_{UV}$, one cannot study correlations much below
that cutoff, so that one cannot follow the coupling into the UV.

Either way, limits on the accuracy of our detectors are not built
into our classical, large-N discussion of the AdS/CFT
correspondence; this is why we have seen no hint of irreversibility
in our discussion. Of course, one could also study the duals of gauge theories
which are explicitly cut off, as in \refs{\HellermanQA,\EvansEQ}.

\newsec{Conclusions}

We have resolved the apparent tension between the first-order RG
equations of a quantum field theory and the second-order
supergravity equations which are supposed to encode the RG flow in
the dual asymptotically-AdS spacetime. The essential point is that
the RG behavior of the field theory is contained in {\it two}
first-order equations -- the Callan-Symanzik equations, and the
equations for the evolution of the couplings.  The former depends on
the choice of quantum state, which is the additional information one
needs to specify the most general solution to the bulk, second-order
supergravity equations.

A number of puzzles remain.  In particular, we would like to better
understand the relationship between the bulk fields and the boundary
coupling deep in the IR, as discussed in \S4.5; and we would like to
understand the apparent modification of the Callan-Symanzik
equations (including the $\{\Gamma,\Gamma\}$ term) in Eq. \genCS.

\bigskip
\centerline{\bf Acknowledgements}

We would like to thank  C. Beasley,  M. Headrick, D. Kabat, H. Liu,
M.S. Sheikh-Jabbari, and J. Terning for helpful conversations. We
would particularly like to thank O. Aharony, J. de Boer, D. Marolf,
and H. Schnitzer for comments on the draft, as well as for very
helpful discussions and correspondence. We would also like to thank
the MIT Center for Theoretical Physics for their hospitality during the
completion of this project.  A.L. would like to thank
D.Z. Freedman and M. Headrick for discussions on related subjects
during the incubation of \refs{\FreedmanWX}. He would also like to
thank the UC Davis cosmology and particle physics groups, the
Stanford theory group, and the IPM string theory group for their
generous hospitality during parts of this project. The research of
A.L. and A.S. is supported in part by NSF grant PHY-0331516, and by
DOE Grant No. DE-FG02-92ER40706 via an Outstanding Junior
Investigator award.

We would like to dedicate this work to the memories of John Brodie
and Andrew Chamblin.

\listrefs
\end